\documentclass[aip,jcp,reprint,amsmath,amssymb,floatfix,citeautoscript,longbiography]{revtex4-2}
\usepackage{cancel}
\usepackage{amsmath}
\usepackage{mathtools}
\usepackage{physics}
\usepackage{graphicx}
\usepackage{amssymb}
\usepackage{amsthm}
\usepackage{bm}
\usepackage{dcolumn}
\usepackage{braket}
\usepackage{longtable}
\usepackage{booktabs}
\usepackage{ragged2e}
\usepackage{newtxtext,newtxmath}
\usepackage[version=3]{mhchem} 
\usepackage[T1]{fontenc}       
\usepackage{pstricks}
\usepackage[colorlinks=true,allcolors=blue]{hyperref}
\usepackage[capitalise]{cleveref}
\usepackage{xcolor}

\let\vr\undefined
\newcommand{\vr}{{\bm{r}}}

\newcommand{\vq}{{\bm{q}}}

\newcommand{\vk}{{\bm{k}}}
\newcommand{\vR}{{\bm{R}}}

\newcommand{\vT}{{\bm{T}}}

\newcommand{\vG}{{\bm{G}}}
\newcommand{\vQ}{{\bm{Q}}}

\newcommand{\vx}{{\bm{x}}}

\DeclareMathOperator\erfc{erfc}

\allowdisplaybreaks

\begin{document}

\title{Wavefunction-based periodic quantum chemistry}

\author{Hong-Zhou Ye}
\email{hzye@umd.edu}
\affiliation{Department of Chemistry and Biochemistry, University of Maryland, College Park, MD, 20742, USA}
\affiliation{Institute for Physical Science and Technology, University of Maryland, College Park, MD, 20742, USA}
\author{Timothy C. Berkelbach}
\email{t.berkelbach@columbia.edu}
\affiliation
{Department of Chemistry, Columbia University, New York, New York 10027, USA}
\affiliation
{Initiative for Computational Catalysis, Flatiron Institute, New York, New York 10010, USA}

\begin{abstract}
High-accuracy molecular quantum chemistry offers a promising toolbox for
applications to condensed-phase systems, but this field is difficult to enter
due to its combination of topics from molecular quantum chemistry, solid-state
physics, and numerical methods.
Aiming to ease this transition, we present a comprehensive, pedagogical
tutorial on periodic quantum chemistry calculations, using both mean-field and
correlated theories.
The subtleties of periodic Coulomb interactions are discussed in detail,
focusing on the Ewald summation approach. 
We describe the two most popular periodic, one-electron basis
functions---plane waves and periodic linear combinations of atomic
orbitals---and we give formulas for all Hamiltonian integrals. 
Next, we explain the use of $k$-point sampling as a symmetry adaptation of
supercell basis functions and the associated savings in computational costs as
well as the use of density fitting and related approximations to further reduce
costs.
We present the working equations of a few example periodic quantum chemistry
methods, including Hartree-Fock theory, perturbation theory, and
coupled-cluster theory, and we discuss their finite-size errors and convergence
to the physically relevant thermodynamic limit.
Finally, we briefly discuss local correlation and quantum embedding theories,
which are especially appropriate for periodic systems due to their
lattice translational symmetries.
\end{abstract}

\maketitle

\section{Introduction}

The accurate simulation of the electronic structure of extended systems beyond
one-electron theories, such as Hartree-Fock (HF) and density functional theory (DFT),
is a longstanding challenge.
Although many efforts exist, one promising direction is the application of the
wavefunction-based methods of computational quantum chemistry, which are commonly used to study
the correlated electronic structure of increasingly large molecules with high
precision.
This idea is an old one, and the goal is of course not to explicitly simulate
$10^{23}$ electrons as if the solid were a large molecule, but rather to
approach this thermodynamic limit via finite systems with periodic boundary
conditions.
However, the apparent computational costs are high due to the need to eliminate
finite-size errors in addition to the usual basis set and electron correlation
errors of molecular quantum chemistry.
These costs have limited the accuracy of early calculations and---perhaps in
combination with some cultural differences between chemistry and physics---have
otherwise discouraged serious, widespread community development.
Some of the most notable pioneering efforts for general-purpose software were
those leading to the CRYSTAL code in the
1980s~\cite{Pisani1980,Dovesi1983,Dovesi2020} and the associated CRYSCOR code in the late
2000s~\cite{Pisani2012}.

In recent years, improved computing resources, algorithmic developments, and
dedicated efforts from an increased number of research groups have enabled
wavefunction-based quantum chemical methods to be applied to periodic systems
with the same level of rigor typical of molecular quantum chemistry. A
non-exhaustive survey of existing software hints at this changing tide:
at the time of writing,
software that implements any kind of correlated theory (i.e., beyond HF or DFT)
with periodic boundary conditions include (alphabetically)
Cc4s~\cite{cc4s},
CP2K~\cite{Kuehne2020},
CRYSTAL/CRYSCOR~\cite{Dovesi2020,Pisani2012},
FHI-aims~\cite{Blum2009},
GPAW~\cite{Mortensen2024},
PySCF~\cite{Sun2017a,Sun2020},
and
VASP~\cite{Hafner2008},
as well as development versions of
Q-Chem~\cite{Lee2021,Lee2022},
and TURBOMOLE~\cite{Balasubramani2020,Franzke2023,Nejad2025,Zhu2025}.
Similarly, traditionally molecular quantum chemistry codes that implement
periodic boundary conditions for mean-field calculations include
Gaussian~\cite{Kudin2000,Izmaylov2006} and a development version of
TeraChem~\cite{Wang2024}.  Within the next decade, we expect that correlated
calculations of periodic systems will be as accessible as those for molecules.

To engage with this field requires knowledge of elements
of both solid-state physics and molecular quantum chemistry,
which can present a high barrier for entry.
With this tutorial, we hope to provide the necessary background for a
relative newcomer as well as a technical reference document for a
computational scientist looking to understand existing software or
develop their own.
Because this is a tutorial article, our references will
prioritize education and recent developments, and not attribution.
We will assume basic knowledge of electronic structure theory and
molecular quantum chemistry, as available in standard
textbooks~\cite{Szabo1996}.
We also refer the reader to several other previously published books and review
articles on closely related
topics~\cite{Pisani1988,Evarestov2012,Pisani1996,Kratzer2019,Zhang2019,Robinson2025}.

Throughout this article, we interchangeably use the imperfect phrases
``wavefunction-based'' or ``quantum chemistry'' methods to refer to ab
initio methods that use a finite basis set to represent orbitals that make up
many-electron wavefunctions.  This includes HF theory,
second-order M{\o}ller-Plesset perturbation theory (MP2), and coupled-cluster (CC) theory---which
might be considered the workhorse methods of single-reference, molecular
quantum chemistry (despite the latter's birth in nuclear
physics~\cite{Coester1960} and early applications to periodic
systems~\cite{Freeman1977})---as well as auxiliary-field quantum Monte
Carlo, multiconfigurational methods, matrix-product state methods,
and more.
Within many-body perturbation theory, the boundary between
wavefunction and Green's function methods is blurry, and many of the latter,
such as the random-phase approximation or the GW approximation, also fit within
the scope of this article. Finally,
there is also overlap with
real-space quantum Monte Carlo, which has been important for the study of correlated
electrons in extended systems for many years~\cite{Foulkes2001},
and density functional theory (DFT),
especially in the context of hybrid and double-hybrid functionals.

The layout of this tutorial roughly mirrors the workflow of a periodic quantum
chemistry calculation.
In Sec.~\ref{sec:ham}, we define the periodic Hamiltonian in first
quantization, within the standard Born-Oppenheimer approximation of electronic
structure theory.
Because most quantum chemical methods are performed using finite basis sets, in
Sec.~\ref{sec:basis}, we introduce the two most common periodic basis
functions---plane-waves and periodic linear combinations of atom-centered orbitals---and define the
needed Hamiltonian integrals over these basis functions.
We also discuss the exploitation of lattice translational symmetry and its
associated conservation of crystal momentum, leading to important computational
savings.
In Sec.~\ref{sec:compression}, we discuss the compression of electron
repulsion integrals over atom-centered basis functions (e.g., using the density fitting
approximation).
In Sec.~\ref{sec:theory}, we present the working equations of a few example
wavefunction theories, including HF, MP2, and CCSD,
as well as a discussion of finite-size errors.

Everything up to this point of the tutorial will constitute ``canonical''
periodic quantum chemistry performed in the delocalized basis of crystalline
orbitals, analogous to molecular quantum chemistry performed in the delocalized
basis of molecular orbitals.  It is well-known that in large molecules, such a
formulation is responsible for an unphysically high scaling of the
computational cost with system size.  Therefore, in Sec.~\ref{sec:embed}, we
briefly discuss periodic local correlation and the closely related embedding
approximations. Finally, in Sec.~\ref{sec:conc}, we conclude with a perspective
and the identification of future research directions.

\section{Periodic Hamiltonian}
\label{sec:ham}

\begin{figure}[t]
	\includegraphics[scale=1.0]{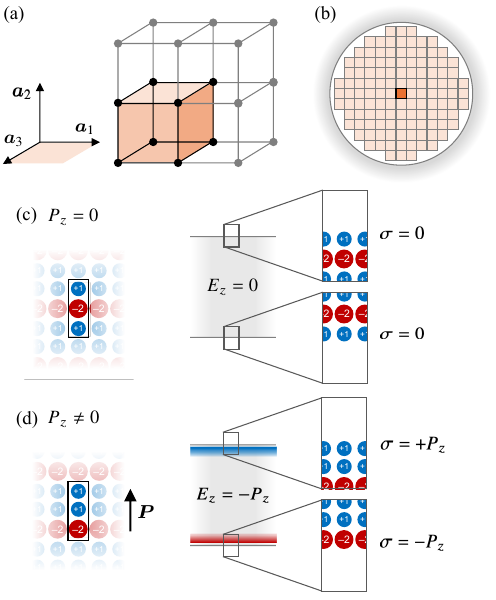}
	\caption{
(a) Example crystal lattice structure, with a single unit cell highlighted.
The unit cell is defined by the primitive lattice vectors $\bm{a}_1$,
$\bm{a}_2$, $\bm{a}_3$ and has volume
$\Omega = |\bm{a}_1\cdot(\bm{a}_2\times\bm{a}_3)|$.
(b) Collection of unit cells, for which the energy of the central cell
depends on the shape (here, a sphere) and the dielectric environment.
The shape dependence is the origin of the conditional convergence
in the definition of the electrostatic potential energy of the
central cell.
(c) A unit cell choice with zero dipole moment generates no surface charge
density, and thus no electric field inside the slab.
(d) A unit cell choice with nonzero dipole moment $\bm{P}$ generates a surface termination
with equal but opposite surface charge densities $\sigma = \pm P_z$, which generates
a constant electric field $E_z = -P_z$ everywhere inside the slab.
}
	\label{fig:lattice}
\end{figure}

Although our understanding of the electronic structure of solids has been undeniably advanced by the study of model Hamiltonians with simplified interactions, here we are interested in a fully ab initio description.
Within the Born-Oppenheimer approximation, the Hamiltonian describes the dynamics of electrons in the fixed field of the nuclei. 
Defining the potential energy, which includes nuclear-nuclear, electron-nuclear, and electron-electron Coulomb interactions, is straightforward for finite systems, but less so for infinite, periodic systems because of the long-range nature of the Coulomb interaction.
The approach we will describe, commonly referred to as Ewald summation, is the same one used to calculate the Coulomb energy in classical periodic systems of charged particles, but it may be unfamiliar to electronic structure theorists.

\subsection{Periodic electrostatics and Ewald summation}

We consider a unit cell defined by the primitive lattice vectors $\bm{a}_1, \bm{a}_2, \bm{a}_3$ with volume $\Omega = |\bm{a}_1\cdot(\bm{a}_2 \times \bm{a}_3)|$.
As shown in Fig.~\ref{fig:lattice}(a), the unit cell is periodically repeated in space according to the infinite set of lattice translation vectors $\vT = n_1 \bm{a}_1 + n_2\bm{a}_2 + n_3\bm{a}_3$, where $n_i$ are integers.  
Throughout this tutorial, we assume the crystal is three-dimensional; many of the same ideas and methods apply in lower dimensions, although the precise formulas will vary, and some aspects of series convergence will differ.
Because Fourier transforms are ubiquitous in the mathematical analysis of periodic functions, let us define our convention for future reference: for a function $f(\vr)$ that is periodic over the volume $\Omega$ generated by the primitive lattice vectors $\bm{a}_1$, $\bm{a}_2$, $\bm{a}_3$,
\begin{subequations}
\label{eq:ft_conv}
\begin{align}
f(\vr) &= \frac{1}{\Omega} \sum_\vG e^{i\vG\cdot\vr} f(\vG), \\
f(\vG) &= \int_\Omega d^3r\ e^{-i\vG\cdot\vr} f(\vr),
\end{align}
\end{subequations}
where $\vG$ is a reciprocal lattice vector,
$\vG = n_1\bm{b}_1 + n_2\bm{b}_2 + n_3\bm{b}_3$,
and $\bm{b}_i$ are primitive reciprocal lattice vectors---e.g., $\bm{b}_1 = (2\pi/\Omega) \bm{a}_2 \times \bm{a}_3$, and cyclic permutations---which satisfy $\bm{a}_i \cdot \bm{b}_j = 2\pi \delta_{ij}$.
Reciprocal lattice vectors satisfy $e^{i\vG\cdot\vT} = 1$.

We focus on a central unit cell inside the crystal, as shown in Fig.~\ref{fig:lattice}(b).
We assume the unit cell is charge neutral and contains $N$ electrons with positions
$\vr_n$ and $M$ nuclei with positions $\vR_A$ and charges $Z_A$.
The kinetic energy of the electrons in the unit cell is
$T_\mathrm{e} = - \frac{1}{2} \sum_{n=1}^{N} \nabla_n^2$.
The Coulomb energy of the central unit cell is
\begin{subequations}
\label{eq:Vnaive}
\begin{align}
V &= \tilde{V}_\mathrm{nn} + \tilde{V}_\mathrm{en} + \tilde{V}_\mathrm{ee}, \\
\tilde{V}_\mathrm{nn} &= \frac{1}{2}\sum_{A,B=1}^{M} \sideset{}{'}\sum_{\vT} \frac{Z_A Z_B}{|\vR_A-\vR_B-\vT|}, \\
\tilde{V}_\mathrm{en} &= -\sum_{n=1}^{N}\sum_{A=1}^{M}\sum_\vT \frac{Z_A}{|\vr_n-\vR_A-\vT|}, \\
\tilde{V}_\mathrm{ee} &= \frac{1}{2}\sum_{n,n'=1 }^{N}\sideset{}{'}\sum_\vT \frac{1}{|\vr_n-\vr_{n'}-\vT|},
\end{align}
\end{subequations}
where the primed summation excludes self-interactions when $\vT=\bm{0}$; except where indicated otherwise, $\sum_{\vT}$ indicates an infinite sum over all lattice translation vectors.  
By including only a single sum over $\vT$, Eq.~(\ref{eq:Vnaive}) is the Coulomb energy of the central unit cell, i.e., it is an intensive quantity.
However, each individual term in Eq.~(\ref{eq:Vnaive}) diverges ($\sum_{\vT\neq \bm{0}}^{\bm{T}_\text{max}} 1/T \sim \int^{\vT_\text{max}}4\pi R^2dR/R \sim \vT_\text{max}^2$).
When all three are treated together, the total sum no longer diverges because the cell is charge neutral, but the sum is only conditionally convergent, meaning that its value depends on the order of summation.

What is the physical origin of this conditional convergence?  For a crystal composed of copies of a neutral unit cell, the electrostatic energy of a central cell depends on the unit cell dipole moment $\bm{P}$,
the crystal shape, and the crystal environment, due to a shape-dependent contribution from the surface, and different summation orderings correspond to different shapes. 
In other words, the contribution of the surface to the energy of a central cell cannot be neglected, no matter how large the crystal is!

A more practical decomposition of the energy of the central cell is
\begin{equation}
\label{eq:V}
V = V_\mathrm{nn} + V_\mathrm{en} + V_\mathrm{ee} + V_\mathrm{surf}
\end{equation}
where the first three terms are shape-independent quantities defined by
convergent sums (described below), and the surface term isolates the shape and
environment dependence.
For example, consider an increasingly large, spherical cluster of unit cells,
as shown in Fig.~\ref{fig:lattice}(b), which corresponds to a lattice summation
order in spherical shells  (i.e., the lattice summations in Eq.~(\ref{eq:Vnaive})
are restricted to $|\vT| < R_\mathrm{c}$ with $R_\mathrm{c} \rightarrow
\infty$). For this shape in vacuum, the surface contribution to the
energy can be shown to be $V_\mathrm{surf} = (2\pi/3\Omega)|\bm{P}|^2$. Other dielectric environments or other shapes
have different surface contributions; for example, when
the collection is surrounded by a perfect conductor
(sometimes called tinfoil boundary conditions), the surface energy vanishes.
To understand the dependence on the unit cell dipole moment $\bm{P}$, see the
two examples in Fig.~\ref{fig:lattice}(c) and (d), which correspond to the same
bulk crystal but different choices of the unit cell. Even after many replicas,
the unit cell with a nonzero dipole moment generates equal and opposite surface charge
densities leading to a nonzero electric field everywhere inside the slab, independent
of the slab thickness.
In practice, we are usually interested in the shape-independent contribution to
the energy, and from now on, we will assume a scenario where the surface
contribution can be neglected ($V_\mathrm{surf}=0$).

The evaluation of the shape-independent, absolutely convergent summations
defining the first three terms of Eq.~(\ref{eq:V}) is most easily
performed using the Ewald summation method, which splits the Coulomb
potential into a short-range (SR) and long-range (LR) component, commonly via the form
\begin{equation}
\label{eq:vr_split}
v(\vr) = v^\mathrm{SR}(\vr) + v^\mathrm{LR}(\vr) 
    = \frac{\mathrm{erfc}(\eta r)}{r} + \frac{\mathrm{erf}(\eta r)}{r}.
\end{equation}
With this splitting, short-range contributions can be evaluated efficiently in
real space, long-range contributions can be evaluated efficiently in reciprocal
space (see below), and $\eta$ is a parameter that can be chosen to balance the
total evaluation cost.

While this formalism provides an efficient procedure for evaluating the total
shape-independent electrostatic energy, 
we might like to assign separate energies to the nuclear
repulsion, the electron-nuclear attraction, and the electron repulsion---as we
do for finite systems---but the nuclei and electrons are not separately charge
neutral, so their individual Coulomb energies are divergent. Therefore, we add
uniform negative and positive charge densities, combining the negative density
with the nuclei and the positive density with the electrons; both systems are
now charge neutral, allowing for an energy assignment. However, it must always
be kept in mind that the nuclear energy (for example) is really the energy of
the nuclei plus a uniform negative charge density.

When all of the above is performed, the total Coulomb energy is given by
Eq.~(\ref{eq:V}), with
\begin{subequations}
\label{eq:Vewald}
\begin{align}
\label{eq:Vnn}
V_\mathrm{nn} &=
    \frac{1}{2}\sum_A Z_A^2 v_\mathrm{M}
    + \frac{1}{2} \sum_{A\neq B} Z_A Z_B v_\mathrm{E}(\vR_A-\vR_B) \\
\label{eq:Ven}
V_\mathrm{en} &= -\sum_{nA} Z_A v_\mathrm{E}(\vr_n-\vR_A) \\
\label{eq:Vee}
V_\mathrm{ee} &= \frac{1}{2} N v_\mathrm{M}
    + \frac{1}{2} \sum_{n\neq n'} v_\mathrm{E}(\vr_n-\vr_{n'}),
\end{align}
\end{subequations}
where the so-called Ewald potential is
\begin{equation}
\label{eq:ewald}
v_\mathrm{E}(\vr) = \frac{4\pi}{\Omega}\sum_{\vG\neq\bm{0}} e^{i\vG\cdot\vr}
    \frac{e^{-G^2/4\eta^2}}{G^2}
    + \sum_\vT \frac{\erfc(\eta|\vr-\vT|)}{|\vr-\vT|}
    - \frac{\pi}{\Omega\eta^2}
\end{equation}
and the generalized Madelung constant is
\begin{equation}
\label{eq:madelung}
\begin{split}
v_\mathrm{M} &= \lim_{\vr\rightarrow 0} \left[v_\mathrm{E}(\vr) - 1/r\right] \\
    &= \frac{4\pi}{\Omega}\sum_{\vG\neq\bm{0}}
    \frac{e^{-G^2/4\eta^2}}{G^2}
    + \sum_{\vT\neq \bm{0}} \frac{\erfc(\eta|\vT|)}{|\vT|}
    -\frac{2\eta}{\sqrt{\pi}} - \frac{\pi}{\Omega\eta^2}.
\end{split}
\end{equation}
(We suppose $v_\mathrm{M}$ is called the Madelung constant because of its similarity to the textbook quantity of the same name, i.e., the potential at ion sites in an ionic crystal.
To the best we can find, this approach to periodic electrostatics, including the names ``Ewald potential'' and ``Madelung constant'' originate with a 1966 Monte Carlo study of the one-component plasma~\cite{Brush1966}, and was subsequently adopted for quantum Monte Carlo.)
A quick derivation of these equations for a general, neutral collection of point charges is given in App.~\ref{app:ewald}, and more rigorous derivations can be found in Refs.~\onlinecite{Redlack1975,deLeeuw1980,Ballenegger2014}.

Physically, the one-body terms in $V_\mathrm{nn}$ and $V_\mathrm{ee}$ 
correspond to the interaction energy of
each charge with its periodic images (correctly excluding the
self-interaction), and the two-body terms correspond to the interaction energy
of each pair of distinct charges, including their images; both terms also
account for the interaction with the uniform neutralizing density.  Evaluated
according to Eqs.~(\ref{eq:Vewald}), all contributions to the potential energy
are well-defined (all infinite sums are absolutely convergent for $\eta>0$) and
independent of $\eta$, which can be chosen to make the real-space and
reciprocal-space summations converge quickly.

\subsection{The Ewald potential}
\label{sec:ewald}

\begin{figure}[t]
	\includegraphics[scale=1.0]{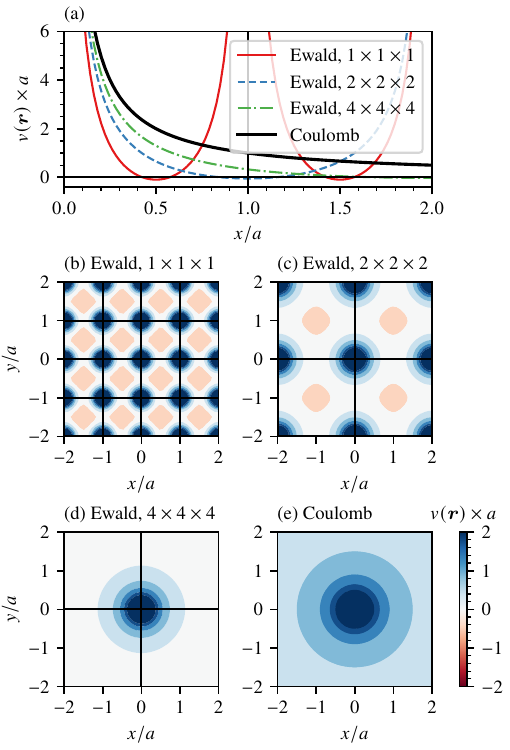}
	\caption{
Plots of the three-dimensional Ewald potential for a cubic cell of side length $a$.
One-dimensional slices are shown in (a) and two-dimensional slices are shown in (b)-(e).
The Ewald potential is plotted for supercells of the sizes indicated along with the
Coulomb potential of a non-periodic system.
}
	\label{fig:ewald}
\end{figure}

The Ewald potential $v_\mathrm{E}(\vr)$ plays an important role in periodic electronic structure theory, replacing the more familiar Coulomb potential.
In our experience, it is a common source of confusion. 
Here we list some of its most important properties.

\begin{itemize}

\item
The Ewald potential is the potential at $\vr$ due to a unit point charge
at the origin,
its periodic images, and its uniform neutralizing density.
The Madelung constant is this potential at the location of the point charge,
excluding its own Coulomb potential.
The Madelung constant is negative and depends only on the lattice shape. For example,
the Madelung constant of a cubic crystal with lattice constant $a$ is
$v_\mathrm{M} = -2.837297479/a$.

\item
The Ewald potential, in contrast to the Coulomb potential, is periodic with
respect to lattice translation vectors, $v_\mathrm{E}(\vr-\vT) =
v_\mathrm{E}(\vr)$, it is not spherically symmetric, and [as defined in
Eq.~(\ref{eq:ewald})] it is zero on average,
$\int_\Omega d^3r\, v_\mathrm{E}(\vr) = 0$.  In Fig.~\ref{fig:ewald}, we plot
one-dimensional and two-dimensional slices of the three-dimensional Ewald
potential for a cubic unit cell, a $2\times 2\times 2$ supercell, and a
$4\times 4\times 4$ supercell.  The periodicity over the supercell is apparent,
and the Ewald potential converges to the Coulomb potential as the supercell
size is increased, as discussed more in the next point.

\item
Adding any constant to the Ewald potential,
$\bar{v}_\mathrm{E}(\vr) = v_\mathrm{E}(\vr) + C$, changes
the total potential energy $V$ via
\begin{equation}
\label{eq:const}
\bar{V} = V + \frac{C}{2} \left( Z - N \right)^2,
\end{equation}
where $Z = \sum_A Z_A$ is the total nuclear charge, and the additional term is zero by neutrality.
Even for non-neutral systems, electric potentials and energies are only defined up to a constant.
In finite systems, this constant is commonly fixed by defining the potential to be zero at infinite separation, $v(r\rightarrow \infty) = 0$. 
In a periodic solid, the particles in the unit cell can never be infinitely separated, and so the same limit is not accessible.
Being zero on average can be understood as the analogous convention to fix the constant in periodic systems. 
The Ewald potential has the useful property that it reduces to the traditional Coulomb potential at fixed $\vr$ in the limit $\Omega \rightarrow \infty$
(this limit holds under the addition of any constant that vanishes in the
$\Omega \rightarrow \infty$ limit.)
The freedom to add a constant without changing the total energy can be exploited to eliminate the atypical one-body terms appearing in $V$, via the modification $\bar{v}_\mathrm{E}(\vr) = v_\mathrm{E}(\vr) - v_\mathrm{M}$.
This will be discussed in Sec.~\ref{sec:theory}, in the context of finite-size errors.

\item
The Ewald potential has the Fourier representation
\begin{subequations}
\begin{align}
\label{eq:ewald_ft}
v_\mathrm{E}(\vr) &= \frac{1}{\Omega} \sum_\vG e^{i\vG\cdot\vr} v_\mathrm{E}(\vG), \\
\label{eq:ewaldG}
v_\mathrm{E}(\vG) &= \begin{cases}
4\pi/G^2 & \vG \neq \bm{0} \\
0 & \vG = \bm{0},
\end{cases}
\end{align}
\end{subequations}
which is another commonly used form, especially in plane-wave-based calculations (see Sec.~\ref{ssec:pw}).  The divergent $\vG=\bm{0}$ term is not dropped simply for numerical convenience, but rather because doing so defines the Ewald potential and its associated conventions and physical implications (zero on average, neutralizing background charge density, and vanishing contribution from the surface). 
For future reference, note that shifting the Ewald potential by a constant $C$ simply modifies the $\vG = \bm{0}$ component,
$\bar{v}_\mathrm{E}(\vG=\bm{0}) = \Omega C$.

\end{itemize}

\subsection{The final Hamiltonian}
\label{ssec:hamiltonian_final}

Returning to the electronic structure problem within the Born-Oppenheimer approximation,
we evaluate $V_\mathrm{nn}$ as a scalar contribution
to the energy $E_\mathrm{nn}$. The electronic Hamiltonian in first quantization is then
\begin{equation}
\label{eq:ham_first}
H = E_\mathrm{nn} + T_\mathrm{e} + V_\mathrm{en} + V_\mathrm{ee}.
\end{equation}
The Hamiltonian only depends explicitly on the degrees of freedom in the cell,
which is consistent with its definition as the energy per unit cell.  The
interactions with the periodic images are implicitly included through the
Madelung constant self-energy and the pairwise Ewald potential.

Here and throughout, we assume the use of the bare electron-nuclear interaction
$-Z/r$, although in practice pseudopotentials (also known as effective core
potentials) are commonly used in periodic calculations.  Pseudopotentials are
used for several practical reasons: they lower the number of electrons, they
can be parameterized to approximately capture relativistic effects, and they
lower the number of plane waves needed by methods that use plane waves as a
basis or for long-range integration (as discussed more below).  For the common
case of pseudopotentials that decay as $-Z_\mathrm{eff}/r$ at long range, the
Ewald summation methods described above are necessary. Additional integral
details arise due to the nonlocal projector form of the short-range part of
most pseudopotentials, but because these are not unique to periodic systems, we
do not discuss them in any detail.  For further information, we refer to a
detailed 2011 review article~\cite{Dolg2011} and references therein.

We emphasize that the Hamiltonian derived here, based on Ewald summation, is only
one of many choices for the definition of the Hamiltonian of periodic systems
with Coulomb interactions.  Any modifications that vanish in the physically
relevant thermodynamic limit are equally valid, and different choices typically
generate different finite-size errors.  One can even use different treatments
of the Coulomb interaction for different integrals or contributions to the
total energy, as commonly done for the nonlocal exchange energy appearing in
Hartree-Fock theory and hybrid
DFT~\cite{Gygi1986,Massidda1993,Paier2005,Broqvist2009,Sundararaman2013}.
Unfortunately, these freedoms make it hard to compare the results of correlated
calculations that have not reached the thermodynamic limit.  An advantage of
the presented formalism is that it uniquely defines a many-body Hamiltonian and
an associated set of integrals in a basis, allowing for exchange between codes
and well-defined comparisons.
A further discussion of these points, including the finite-size errors in the
total energy, will be provided in Sec.~\ref{sec:theory}.

\section{Periodic basis functions and integrals}
\label{sec:basis}

The previous section defined the Hamiltonian operator in first quantization,
which is used directly in real-space methods such as variational or diffusion
quantum Monte Carlo~\cite{Fraser1996,Chiesa2006,Drummond2008}.  The transition
to second quantization, as more commonly used in quantum chemistry, requires
the introduction of a set of basis functions $\phi_b(\vr)$ and the evaluation
of integrals of the one- and two-body operators appearing in the Hamiltonian.
Periodic crystals are most commonly simulated with basis functions satisfying
Born-von Karman boundary conditions,
\begin{equation}
\phi_b(\vr+\vT) = \phi_b(\vr)
\end{equation}
(more generally, so-called twisted boundary conditions satisfying $\phi_b(\vr+\vT) = e^{i\vk_\mathrm{s}\cdot\vT}\phi_b(\vr)$ can be useful).
Integration is restricted to the cell with volume $\Omega$.
The crystalline orbitals (eigenfunctions of some one-electron Hamiltonian, analogous to molecular orbitals in finite systems) are then expanded in this basis,
\begin{equation}
\psi_p(\vr) = \sum_b C_{bp} \phi_b(\vr).
\end{equation}

In second quantization, the electronic structure problem is fully
specified by the overlap integrals
\begin{equation}
S_{b_1b_2} = \int_\Omega d^3r\ \phi_{b_1}^*(\vr) \phi_{b_2}(\vr)
\end{equation}
and the Hamiltonian integrals
\begin{align}
    &T_{b_1b_2} = \int_\Omega d^3r\ \phi_{b_1}^*(\vr) \left[ -\tfrac{1}{2} \nabla_\vr^2 \right] \phi_{b_2}(\vr) \\
    &V_{b_1b_2} = -\sum_A Z_A \int_\Omega d^3r\ \phi_{b_1}^*(\vr) v_\mathrm{E}(\vr-\vR_A) \phi_{b_2}(\vr) \\
\begin{split}
&( b_1 b_2 | b_3 b_4 ) = \int_\Omega d^3r_1 \int_\Omega d^3r_2\ \phi_{b_1}^*(\vr_1) \phi_{b_2}(\vr_1) \\
&\hspace{8em} \times v_\mathrm{E}(\vr_1-\vr_2) \phi_{b_3}^*(\vr_2) \phi_{b_4}(\vr_2).
\end{split}
\end{align}
In principle, once a set of basis functions is selected and the above integrals
are evaluated, the problem is fully defined, and any wavefunction-based quantum
chemistry method can be applied.

\subsection{Basis functions}
\label{ssec:basis}

The two most popular periodic basis functions
are plane waves (PWs) and periodic atom-centered basis functions.
PW basis functions (as used in VASP and Cc4s) are
\begin{equation}
\label{eq:pw}
\phi_\vG(\vr) = \frac{1}{\sqrt{\Omega}} e^{i\vG\cdot\vr},
\end{equation}
where the reciprocal lattice vector $\vG$ is also known as the PW momentum vector.
Note that when these PWs are used to expand the crystalline orbitals,
\begin{equation}
\label{eq:pw_mo_expansion}
\psi_p(\vr)
= \sum_\vG C_{\vG p} \phi_\vG(\vr)
= \frac{1}{\sqrt{\Omega}} \sum_\vG C_{\vG p} e^{i\vG\cdot\vr},
\end{equation}
the expansion coefficients are simply proportional to the Fourier transforms [recall
our convention in Eq.~(\ref{eq:ft_conv})],
\begin{equation}
C_{\vG p} = \frac{1}{\sqrt{\Omega}} \int_\Omega d^3r\ \psi_p(\vr) e^{-i\vG\cdot\vr}
    \equiv \frac{1}{\sqrt{\Omega}} \psi_p(\vG).
\end{equation}
Periodic atom-centered basis functions are
\begin{equation}
\label{eq:gto}
\phi_\mu(\vr) = \sum_\vT \tilde{\phi}_\mu(\vr-\vT)
\end{equation}
where $\tilde{\phi}_\mu(\vr)$ is a non-periodic atom-centered basis function, such as a
Gaussian-type orbital,
\begin{equation}
\tilde{\phi}_\mu(\vr) = N_\mu (x-x_\mu)^l (y-y_\mu)^m (z-z_\mu)^n e^{-\alpha |\vr-\vR_\mu|^2},
\end{equation}
or contractions of several such orbitals (as used in CRYSTAL/CRYSCOR, CP2K, and PySCF),
where $l+m+n$ is the total angular momentum,
the exponent $\alpha$ defines the diffuseness,
and $\vR_\mu=(x_\mu, y_\mu, z_\mu)$ is the position of the atom in the unit cell 
on which the orbital is centered.
Closely related to Gaussian-type orbitals are numerical atomic orbitals (as used in FHI-aims);
we will henceforth refer to all atom-centered basis functions as atomic
orbitals (AOs).

The number of basis functions used in a calculation are chosen differently.
For PWs, a common choice is to limit the number of PWs by a kinetic energy cutoff,
$G^2/2 < E_\mathrm{cut}$, which can be increased towards convergence~\cite{Shepherd2012}.
For periodic AOs, typical quantum chemistry basis set hierarchies are used
(e.g., split-valence basis sets with optional polarization and diffuse functions).
In the remainder of this article, we will use $N_\mathrm{PW}$ and $N_\mathrm{AO}$ to indicate
the number of PW or AO basis functions.

Two advantages of PW basis functions are that they are mutually orthogonal
and that the complete basis set limit is straightforwardly approached by
increasing $E_\mathrm{cut}$. By contrast, periodic AOs are nonorthogonal and
require more careful convergence to the complete basis set limit.
Unfortunately, most standard molecular AO basis sets contain AOs that are too diffuse
for use in solids (e.g., exponents $\alpha < 0.1$), due to severe linear dependencies.
Reducing their diffuseness while maintaining high accuracy and transferablity,
for both mean-field and correlated calculations, is a challenge. Example attempts
include the MOLOPT basis sets (originally developed for CP2K)~\cite{VandeVondele2007},
the pob-TZVP basis sets (originally developed for CRYSTAL)~\cite{Peintinger2012},
the unc-def2 basis sets (originally developed for Q-Chem)\cite{Lee21JCP},
and the GTH-cc-pVXZ basis sets (originally developed for PySCF)~\cite{Ye2022}.
In addition to these general-purpose solid-state basis sets, one can also
optimize basis sets for specific materials of
interest~\cite{Daga2020,Zhou21JCTC,Morales20JCP,Dong24PRB}, although this
could result in reduced transferability.~\cite{Daga2020}
On the other hand, the advantage of AOs is that a much smaller number
is typically required for convergence, especially for systems with non-uniform
electron densities. Moreover, core orbitals can
be straightforwardly treated with AOs, whereas their treatment with PWs
requires impractically high kinetic energy cutoffs or projector augmented
wave techniques~\cite{Bloechl1994,Rostgaard2009}.
For this reason, PWs are commonly used with pseudopotentials that replace the core orbitals and lead to slowly varying valence orbitals with fewer nodes.

\subsection{PW integrals}
\label{ssec:pw}

\begin{figure}
	\includegraphics[scale=1.0]{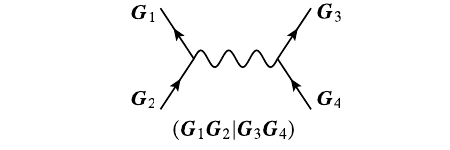}
	\caption{
    The Coulomb vertex for PW basis functions. Electron 1 (left) with momentum $\vG_2$ interacts with
electron 2 (right) with momentum $\vG_4$, after which electron 1 has momentum $\vG_1$ and electron 2 has
momentum $\vG_3$. The total momentum before and after the interaction must be the same, i.e., $\vG_1+\vG_3 = \vG_2+\vG_4$,
and the magnitude of the momentum transferred is $|\vG_1-\vG_2| = |\vG_3-\vG_4|$.
}
	\label{fig:coulomb}
\end{figure}

Another advantage of PW basis functions over periodic AO basis functions is that
integrals are typically simpler to compute and more efficient to store and manipulate.
The PW basis functions defined in Eq.~(\ref{eq:pw}) are orthonormal, so the
overlap matrix is the identity matrix,
\begin{equation}
S_{\vG\vG'} = \frac{1}{\Omega} \int_\Omega d^3r\ e^{-i(\vG-\vG')\cdot\vr} = \delta_{\vG\vG'}.
\end{equation}
The kinetic energy matrix of integrals is also diagonal,
\begin{equation}
T_{\vG\vG'} = \frac{1}{\Omega}
    \int_\Omega d^3r\ e^{-i\vG\cdot\vr} \left[-\tfrac{1}{2}\nabla_\vr^2\right] e^{i\vG'\cdot\vr}
    = \frac{G^2}{2} \delta_{\vG\vG'}.
\end{equation}
The electron-nuclear integrals are
\begin{equation}
\begin{split}
\label{eq:elnuc_pw}
V_{\vG\vG'} &= -\frac{1}{\Omega}\sum_A Z_A \int_\Omega d^3r\ e^{-i(\vG-\vG')\cdot\vr}
    v_\mathrm{E}(\vr-\vR_A) \\
    &= -\Omega^{-1} v_\mathrm{E}(\vG-\vG')\rho_\mathrm{n}(\vG-\vG')
\end{split}
\end{equation}
where $v_\mathrm{E}(\vG)$ is the Fourier transform of the Ewald potential [Eq.~(\ref{eq:ewaldG})],
and $\rho_\mathrm{n}(\vG) = \sum_A Z_A e^{-i\vG\cdot\vR_A}$ is the Fourier transform of the
nuclear charge density (sometimes called the structure factor).
Finally, the two-body electron repulsion integrals (ERIs) are
\begin{equation}
\label{eq:pw_eri}
\begin{split}
( \vG_1\vG_2|\vG_3\vG_4)
    &= \frac{1}{\Omega^2} \int_\Omega d^3r_1 \int_\Omega d^3 r_2\ e^{i\vG_{12}\cdot\vr_1} \\
    &\hspace{3em} \times v_\mathrm{E}(\vr_1-\vr_2) e^{i\vG_{34}\cdot\vr_2} \\
 &= \Omega^{-1} v_\mathrm{E}(\vG_{34}) \delta_{\vG_1+\vG_3,\vG_2+\vG_4},
\end{split}
\end{equation}
where $\vG_{12} = \vG_2-\vG_1$.
The delta function expresses the conservation of momentum by the Ewald
interaction, as shown in Fig.~\ref{fig:coulomb}.
This conservation law arises because the interaction depends only on the difference
of particle positions, as discussed more in Secs.~\ref{ssec:kpts} and \ref{ssec:kbasis}.
Note that despite the large formal size of the ERI tensor, which is $O(N_\mathrm{PW}^4)$,
its elements are simple and can be cheaply recomputed as needed or
stored in a one-dimensional $O(N_\mathrm{PW})$ array. This is a major advantage of PWs over AOs.

\subsection{Periodic AO integrals}
\label{ssec:gto}

Evaluating and manipulating periodic AO integrals is more complicated. In principle,
they can be reduced to infinite lattice summations of non-periodic AO integrals.
For example, the overlap integral is
\begin{equation}
\label{eq:Smn}
\begin{split}
S_{\mu\nu} &= \int_\Omega d^3r\ \phi_\mu^*(\vr) \phi_\nu(\vr) \\
&= \sum_{\vT\vT'} \int_\Omega d^3r\ \tilde{\phi}_\mu^*(\vr-\vT) \tilde{\phi}_\nu(\vr-\vT') \\
&= \sum_{\vT} \int d^3r\ \tilde{\phi}_\mu^*(\vr) \tilde{\phi}_\nu(\vr-\vT)
\equiv \sum_{\vT} S_{\mu\nu}^{\bm{0}\vT}
\end{split}
\end{equation}
where $S_{\mu\nu}^{\bm{0}\vT}$ is a non-periodic AO integral, which is available in any quantum chemistry package.
Note that one of the lattice summations is used to extend the domain of integration to all space.
The kinetic energy integral can be defined in the same way,
\begin{equation}
\label{eq:Tmn}
T_{\mu\nu} = \sum_\vT \int d^3r\
    \tilde{\phi}_\mu^*(\vr)\left[-\tfrac{1}{2}\nabla_\vr^2\right]\tilde{\phi}_\nu(\vr-\vT)
    = \sum_\vT T_{\mu\nu}^{\bm{0}\vT}.
\end{equation}
Both the overlap and kinetic energy integrals can be
efficiently evaluated directly by lattice summation, as written, due
to the exponential decay of the matrix elements with the separation $\vT$.

The one- and two-body Coulomb integrals are
\begin{align}
\begin{split}
&V_{\mu\nu} = -\sum_{\vT A} Z_A \int d^3r\ \tilde{\phi}_\mu^*(\vr) v_\mathrm{E}(\vr-\vR_A) \tilde{\phi}_\nu(\vr-\vT) \\
    &\hspace{2em} \equiv \sum_{\vT} V_{\mu\nu}^{\bm{0}\bm{T}}
\end{split}
\\
\label{eq:eris}
&(\mu\nu|\kappa\lambda) = \sum_{\vT_1\vT_2\vT_3} (\mu^{\bm{0}} \nu^{\vT_1}|\kappa^{\vT_2} \lambda^{\vT_3})
\end{align}
where non-periodic integrals are calculated using the Ewald potential in Eq.~(\ref{eq:ewald}).
First, consider the one-body Coulomb integrals. The integrals decay
exponentially with $\vT$ due to the orbital overlap, but---due to the analytic form of
the Ewald potential---it is most practical to separately evaluate the
SR, LR, and constant contributions,
\begin{equation}
\label{eq:Vmn}
V_{\mu\nu} = V_{\mu\nu}^\mathrm{SR} + V_{\mu\nu}^\mathrm{LR} + \frac{\pi}{\Omega \eta^2} Z S_{\mu\nu}.
\end{equation}
For the SR contribution, we have
\begin{equation}
\label{eq:v_sr}
\begin{split}
&V_{\mu\nu}^\mathrm{SR} = -\sum_A Z_A \\
&\hspace{1em}\times \sum_{\vT\vT'} \int d^3r \tilde{\phi}^*_\mu(\vr) \frac{\mathrm{erfc}(\eta|\vr-\vR_A-\vT|)}{|\vr-\vR_A-\vT|}
    \tilde{\phi}_\nu(\vr-\vT'),
\end{split}
\end{equation}
which is a rapidly converging lattice summation.
The LR contribution can be evaluated using the Fourier transform of the orbital pair density,
\begin{subequations}
\label{eq:pwdf}
\begin{align}
\rho_{\mu\nu}(\vr) &= \frac{1}{\Omega} \sum_\vG \rho_{\mu\nu}(\vG) e^{i\vG\cdot\vr} \\
\rho_{\mu\nu}(\vG) &= \int_\Omega d^3r\ e^{-i\vG\cdot\vr} \phi_\mu^*(\vr) \phi_\nu(\vr),
\end{align}
\end{subequations}
which can be calculated analytically for Gaussian-type AOs~\cite{Sun2017} or numerically using the fast
Fourier transform, leading to
\begin{equation}
V_{\mu\nu}^{\mathrm{LR}} = -\frac{4\pi}{\Omega}
    \sum_{\vG\neq 0} \rho_{\mu\nu}(-\vG) \frac{e^{-G^2/4\eta^2}}{G^2} \rho_\mathrm{n}(\vG).
\end{equation}

For the two-electron ERIs, the $\vT_1$ and $\vT_3$ lattice summations again converge quickly,
but the $\vT_2$ lattice summation converges slowly. For both practicality and efficiency,
the natural evaluation is
\begin{equation}
\label{eq:eri_ewald}
(\mu\nu|\kappa\lambda) = (\mu\nu|\kappa\lambda)^\mathrm{SR} + (\mu\nu|\kappa\lambda)^\mathrm{LR}
    - \frac{\pi}{\Omega \eta^2} S_{\mu\nu} S_{\kappa\lambda}.
\end{equation}
with
\begin{subequations}
\begin{align}
&(\mu\nu|\kappa\lambda)^\mathrm{SR} = \sum_{\vT_1\vT_2\vT_3} ( \mu^{\bm{0}} \nu^{\vT_1} | \kappa^{\vT_2} \lambda^{\vT_3} )^\mathrm{SR} \\
\label{eq:eri_sr}
\begin{split}
&( \mu^{\bm{0}} \nu^{\vT_1} | \kappa^{\vT_2} \lambda^{\vT_3} )^\mathrm{SR}
    = \iint d^3r_1 d^3 r_2\ \tilde{\phi}_\mu^*(\vr_1) \tilde{\phi}_\nu(\vr_1-\vT_1) \\
    &\hspace{4em} \times \frac{\erfc(\eta r_{12})}{r_{12}} \tilde{\phi}_\kappa^*(\vr_2-\vT_2) \tilde{\phi}_\lambda(\vr_2-\vT_3).
\end{split}
\end{align}
\end{subequations}
and
\begin{equation}
(\mu\nu|\kappa\lambda)^\mathrm{LR} = \frac{4\pi}{\Omega} \sum_{\vG\neq 0}
    \rho_{\mu\nu}(-\vG) \frac{e^{-G^2/4\eta^2}}{G^2} \rho_{\kappa\lambda}(\vG).
\end{equation}

The range separation parameter $\eta$ can be chosen to balance
the number of lattice translation vectors used for evaluating the SR contributions
and
the number of PWs used for evaluating the LR contributions~\cite{Sun2023}.
Non-periodic SR integrals of the types in Eqs.~(\ref{eq:v_sr}) and
(\ref{eq:eri_sr}) are available in many quantum chemistry packages---e.g., due to the
popularity of range-separated hybrid functionals.

In the limit $\eta\rightarrow\infty$, the Ewald integrals can be evaluated entirely in reprocal space,
\begin{align}
V_{\mu\nu} &= -\frac{1}{\Omega} \sum_{\vG} \rho_{\mu\nu}(-\vG) v_\mathrm{E}(\vG) \rho_\mathrm{n}(\vG) \\
\label{eq:eri_gpw}
(\mu\nu|\kappa\lambda) &= \frac{1}{\Omega} \sum_\vG \rho_{\mu\nu}(-\vG)v_\mathrm{E}(\vG) \rho_{\kappa\lambda}(\vG).
\end{align}
One can choose to evaluate the overlap and kinetic energy integrals entirely in
reciprocal space as well,
\begin{align}
S_{\mu\nu} &= \frac{1}{\Omega} \sum_{\vG} \phi_\mu^*(\vG) \phi_\nu(\vG) \\
T_{\mu\nu} &= \frac{1}{\Omega} \sum_{\vG}  \phi_\mu^*(\vG) \tfrac{1}{2} G^2 \phi_\nu(\vG).
\end{align}
This approach was popularized as the Gaussian and plane wave (GPW) method in
the CP2K software package~\cite{Lippert1997,VandeVondele2005}.
The Fourier transform is performed numerically by a
real-space discretization of the orbital pair density onto a grid of $N_\mathrm{PW}$ points
and the fast Fourier transform
algorithm; the densities $\rho_{\mu\nu}(\vG)$ can be obtained with
$O(N_\mathrm{AO}^2 N_\mathrm{PW} \ln N_\mathrm{PW})$ operations and $O(N_\mathrm{AO}^2 N_\mathrm{PW})$ storage costs.
Typically, one requires enough grid points (equivalently, enough PWs)
to accurately resolve the orbitals or pair densities,
requiring the use of pseudopotentials and basis sets without compact AOs.
The Gaussian and augmented plane-wave method represents one possible extension for
all-electron calculations~\cite{Lippert1999}.

Finally, we note that AO integral evaluation is more expensive
for solids than molecules, even for the same number of AO basis functions, due
to overheads associated with lattice summations and PW-based Fourier transforms. 

\subsection{Periodic AO integrals by real-space lattice summation of the Coulomb potential}

A potential disadvantage of the approach presented in the previous subsection \ref{ssec:gto} is its use of PWs and Fourier transforms, which are not common in molecular quantum chemistry packages. 
Many techniques exist to obtain the same results by evaluating only real-space integrals of the standard Coulomb potential, taking care to avoid (or correct) the divergences or conditional convergences described in our presentation of Ewald summation~\cite{Pisani1980,Dovesi1983,HammesSchiffer1994}. 
For example, this latter approach is used in the CRYSTAL/CRYSCOR and TURBOMOLE packages.

In general, the naive expression,
\begin{equation}
(\mu\nu|r_{12}^{-1}|\kappa\lambda) = \sum_{\vT_1\vT_2\vT_3}
    (\mu^{\bm{0}} \nu^{\vT_1}| r_{12}^{-1} | \kappa^{\vT_2} \lambda^{\vT_3})
\end{equation}
evaluated using the Coulomb potential, is not absolutely convergent.
Assume that the pair densities $\rho_{\mu^{\bm{0}} \nu^{\vT_1}}(\vr)$ and $\rho_{\kappa^{\vT_2} \lambda^{\vT_3}}(\vr)$ have lowest nonzero multipoles of order $l_1$ and $l_2$.
Then absolute convergence of the lattice sum requires that $l_1 + l_2 > 2$ (the case $l_1 + l_2 = 2$ is conditionally convergent).
Therefore, rather than treat the integrals as standalone quantities, most real-space methods treat them on a case-by-case basis depending on their contributions to the total energy, Fock matrix, etc. 
For example, the exchange contribution to the Fock matrix
$K_{\mu\nu} = \sum_{\vT_1} K_{\mu\nu}^{\bm{0}\vT_1}$, with
\begin{equation}
K_{\mu\nu}^{\bm{0}\vT_1}
    = \sum_{\kappa\lambda} \sum_{\vT_2\vT_3} (\mu^{\bm{0}} \lambda^{\vT_3} | r_{12}^{-1} | \kappa^{\vT_2} \nu^{\vT_1})
        D_{\lambda\kappa}^{\vT_3 \vT_2},
\end{equation}
where $D_{\lambda\kappa}^{\vT_3\vT_2}$ is the density matrix,
can be safely evaluated and truncated in real space due to the decay of the density matrix with $|\vT_3-\vT_2|$.  
By contrast, the Coulomb contributions from electrons and nuclei have to be treated together to ensure charge neutrality,
\begin{equation}
\label{eq:VJ}
V_{\mu\nu}^{\bm{0}\vT_1} + J_{\mu\nu}^{\bm{0}\vT_1} = \iint d^3r_1 d^3r_2\ \rho_{\mu\nu}^{\bm{0}\vT_1}(\vr_1) r_{12}^{-1}
    \left[\rho_\mathrm{nu}(\vr_2) + \rho_\mathrm{el}(\vr_2)\right].
\end{equation}
Expanding the densities yields an infinite lattice summation; the short-range contributions can be evaluated directly, and the long-range contributions can be evaluated using multipole expansions and classical Ewald-type summations.
Importantly, these approaches do not require Fourier transforms of orbitals or densities.
As discussed at the end of Sec.~\ref{ssec:hamiltonian_final}, the downside of these approaches is that they do not define a Hamiltonian in terms of one- and two-electron integrals, making it difficult to compare methods or codes away from the thermodynamic limit.

\subsection{Supercell, lattice translational symmetry, and k-points}
\label{ssec:kpts}

Results from calculations carried out on a single unit cell using the above Hamiltonian and basis functions will exhibit a finite-size error (see Sec.~\ref{ssec:finite}).
To reduce this error, the simulation cell must be grown to include multiple unit cells (called a ``supercell'').
When doing so, the Ewald potential and the Madelung constant appearing in the many-body Hamiltonian will change (because they depend on the size of the cell through the lattice vectors $\vT$).
Furthermore, more electrons will be explicitly simulated, which increases the computational cost.
For example, consider a periodic HF calculation, whose cost formally increases as $N^4$.
Increasing from a unit cell to a $2\times 2\times 2$ supercell would increase the computational cost by a factor of $8^4 = 4096$; for correlated methods, such as MP2 or CC theory, the increase is even worse.
However, the supercell Hamiltonian has lattice translation symmetries, i.e., it is unchanged when all atoms (or, equivalently, all electrons) are shifted by any of the \textit{unit cell} lattice translation vectors $\vT$, with periodic boundary conditions enforced at the supercell boundaries. 
As usual, these symmetries imply a conservation law, which can be exploited to lower the computational cost.

Specifically, the exploitation of symmetry in electronic structure theory requires
the use of symmetry-adapted basis functions, i.e., ones that transform as the
irreducible representations of the symmetry group. For a periodic supercell
with lattice translation symmetry, the symmetry adapted basis functions
must be eigenfunctions of the lattice translation operator. We describe the determination
and use of these eigenfunctions below, but most of the material is covered
in standard textbooks~\cite{Ashcroft1976,Tinkham2003}.

For notational simplicity, first consider a one-dimensional lattice
with a unit cell length $a$ and a supercell composed of $N_k$ unit cells,
such that the supercell has length $N_ka$.
The symmetry group is defined by the lattice translation
operator $\mathcal{T}(a)$, which generates multiple lattice
translations $\mathcal{T}(na) = \mathcal{T}^n(a)$ with $n<N_k$.
The action of the lattice translation operator on one-electron
functions is $\mathcal{T}(a)f(x) = f(x-a)$.
We can write the one-electron eigenfunctions of the lattice translation operator
as $\psi_k(x)$, where
\begin{equation}    \label{eq:lat_trans_eigenvalue_1d}
\mathcal{T}(a)\psi_k(x) = \psi_k(x-a) = e^{-ika} \psi_k(x).
\end{equation}
By analogy with the
eigenvalues of the \textit{continuous}
translation operator that is generated by the momentum operator,
the parameter $k$ is called the crystal momentum (but there is no crystal momentum operator!).
The eigenvalue of the translation operator must be a phase factor
because the group is cyclic:
performing $N_k$ translations demands that $e^{-iN_kka} = 1$,
and therefore $k = 2\pi n/(N_ka)$, with $n$ an integer.
Because the eigenvalues are identical for $k$ and $k + 2\pi/a$ (equivalently $n$ and $n+N_k$),
it is conventional
to choose the $N_k$ unique values
within the interval $-\pi/a$ to $\pi/a$, known as the first Brillouin zone.
The number of unique crystal momenta $k$ is clearly equal to the number of unit
cells in the supercell $N_k$.
In three dimensions, this argument generalizes to describe supercells
composed of $N_k = N_{k_1}\times N_{k_2} \times N_{k_3}$ unit cells,
\begin{equation}
\label{eq:lat_trans_eigenvalue_3d}
\mathcal{T}(\vT)\psi_\vk(\vr) = \psi_\vk(\vr-\vT) = e^{-i\vk\cdot\vT} \psi_\vk(\vr),
\end{equation}
where $\vT$ is a lattice translation vector of the unit cell and $\vk$ is a
crystal momentum vector in the first Brillouin zone.
Note that $\psi_\vk(\vr)$ and $\psi_{\vk+\vG}(\vr)$ are identical because $\vG\cdot\vT = 1$.
Eigenstates of the lattice translation operator are also commonly written in
the form
\begin{equation}
\label{eq:bloch}
\psi_\vk(\vr) = e^{i\vk\cdot\vr} u_\vk(\vr),
\end{equation}
where $u_\vk(\vr)$ has the periodicity of the unit cell, i.e.,
$u_\vk(\vr-\vT) = u_\vk(\vr)$.
This is the form frequently encountered in statements of Bloch's theorem.

For a one-electron Hamiltonian with lattice translation symmetry, as arises
in mean-field theories, the eigenfunctions can be labeled by the crystal momentum
$\bm{k}$, which is conserved by the Hamiltonian. Therefore, it is natural
to express these eigenfunctions $\psi_{p\vk}(\vr)$ in a basis of functions with the
same symmetry $\phi_{b\vk}(\vr)$,
\begin{equation}
\psi_{p\vk}(\vr) = \sum_{b} C_{bp}(\vk) \phi_{b\vk}(\vr).
\end{equation}
For an $N$-electron Hamiltonian with lattice translation symmetry, the relevant
operator translates all electrons,
\begin{equation}
\mathcal{T}(\vT) = \mathcal{T}_1(\vT) \mathcal{T}_2(\vT) \ldots \mathcal{T}_N(\vT),
\end{equation}
and the symmetry-adapted $N$-electron basis of Slater determinants satisfies
\begin{equation}
\begin{split}
&\mathcal{T}(\vT) |\psi_{1\vk_1}(\vr_1) \psi_{2\vk_2}(\vr_2) \ldots \psi_{N\vk_N}(\vr_N)| \\
    &\hspace{1em} = e^{-i(\vk_1+\vk_2+\ldots\vk_N)\cdot\vT} |\psi_{1\vk_1}(\vr_1) \psi_{2\vk_2}(\vr_2) \ldots \psi_{N\vk_N}(\vr_N)|,
\end{split}
\end{equation}
where $|\cdot|$ is shorthand notation for the Slater determinant.
Therefore, we might guess that the total momentum $\sum_n \vk_n$ is conserved by the $N$-electron Hamiltonian, but this is not quite right.
In general, the sum of several crystal momenta can lie outside the first Brillouin zone, but $\vk$ and $\vk+\vG$ behave identically.
So the total momentum $\sum_n \vk_n$ is conserved up to any reciprocal lattice vector $\vG$, i.e.
\begin{equation}
\sum_{n}\vk_n^{(\text{out})} = \sum_n \vk_n^{(\text{in})} + \vG.   
\end{equation}

The crystal momenta $\vk$ are commonly called $k$-points, and the exact choice of
$N_k$ $k$-points within the first Brillouin zone defines the $k$-point ``mesh''.
The uniform $k$-point meshes derived above are a special case of the Monkhorst-Pack
meshes~\cite{Monkhorst1976} (special because they always include $\vk=\bm{0}$).
In the thermodynamic limit of an infinitely large supercell (and thus $N_k
\rightarrow \infty$), sums become integrals over the Brillouin zone,
\begin{equation}
\frac{1}{N_k} \sum_\vk \rightarrow \frac{\Omega}{(2\pi)^3} \int d^3k
\end{equation}
where $\Omega$ is the volume of the unit cell.
From this point of view, the use of a finite, uniform $k$-point mesh can be seen as an
approximate numerical integration in the style of Riemann summation with $N_k$ grid points,
each associated with volume $(2\pi)^3/(N_k\Omega)$.
Alternative $k$-point meshes that are nonuniform or do not include $\vk=0$ can
thus be justified, even though they do not represent a complete set of symmetry-adapted supercell basis functions (see below).
Finite-size errors can often be analyzed from this integration perspective,
i.e., as the
quadrature error associated with uniform discretization of the Brillouin zone,
which depends on the analytic properties of the integrand.
For example,
if the integrand is analytic and periodic over the Brillouin zone, the integration
error decreases exponentially, as $O(e^{-aN_k})$ (see Refs.~\onlinecite{Weideman2002} and~\onlinecite{Trefethen2014} for pedagogical discussions of this property).
This favorable scaling of the finite-size error is exhibited by semilocal DFT.  
In other cases---e.g., if the integrand
contains an integrable divergence---the scaling of the integration error
is slower and commonly a power law, $O(N_k^{-p})$.

When performing calculations with $k$-points, we commonly need Fourier transforms of functions that are periodic over supercells. 
Keeping $\vG$ as a reciprocal lattice vector of the unit cell, such that $\vk+\vG$ is a reciprocal lattice vector of the super cell, the generalization of our convention in Eq.~(\ref{eq:ft_conv}) is 
\begin{subequations}
\label{eq:ft_k}
\begin{align}
f(\vr) &= \frac{1}{N_k\Omega} \sum_{\vk\vG} e^{i(\vk+\vG)\cdot\vr} f(\vk+\vG) \\
f(\vk+\vG) &=  \int_{N_k\Omega} d^3r e^{-i(\vk+\vG)\cdot\vr} f(\vr),
\end{align}
\end{subequations}
where $N_k\Omega$ indicates integration over the supercell volume.
For eigenfunctions of the lattice translation operator, this simplifies to
\begin{subequations}
\label{eq:ft_k_sym}
\begin{align}
f_\vk(\vr) &= \frac{1}{N_k\Omega} \sum_{\vG} e^{i(\vk+\vG)\cdot\vr} f_\vk(\vk+\vG) \\
f_\vk(\vk+\vG) &=  N_k\int_\Omega d^3r e^{-i(\vk+\vG)\cdot\vr} f_\vk(\vr).
\end{align}
\end{subequations}
We note that some papers and codes, including PySCF, do not always include factors of $N_k$ in their Fourier transform conventions.

Before presenting basis functions with $k$-points and their associated
integrals, we pause to emphasize that this exploitation of lattice translation
symmetry is entirely analogous to the use of point group symmetry in molecular
quantum chemistry calculations: by using a symmetry-adapted set of atomic or
molecular orbitals that each transform as an irreducible representation of the
group, we can immediately identify integrals that vanish by symmetry, leading
to savings in both storage and computations.
Specifically, the Hamiltonian integrals vanish unless the direct product of
the irreducible representations of the orbitals transforms as the totally symmetric
irreducible representation.
Mean-field and correlated calculations in a basis of symmetry-adapted
orbitals that transform as irreducible representations of an Abelian
group are well understood and widely implemented in molecular quantum chemistry;
for example, see Ref.~\onlinecite{Stanton1991}.
Periodic calculations with $k$-points are no different.
In fact, periodic crystals are categorized by their space group symmetry, which
is the combination of point group and translation group symmetries.
Consideration of space group symmetry can be used to restrict attention
to the so-called irreducible Brillouin zone.

\begin{figure*}
	\includegraphics[scale=1.0]{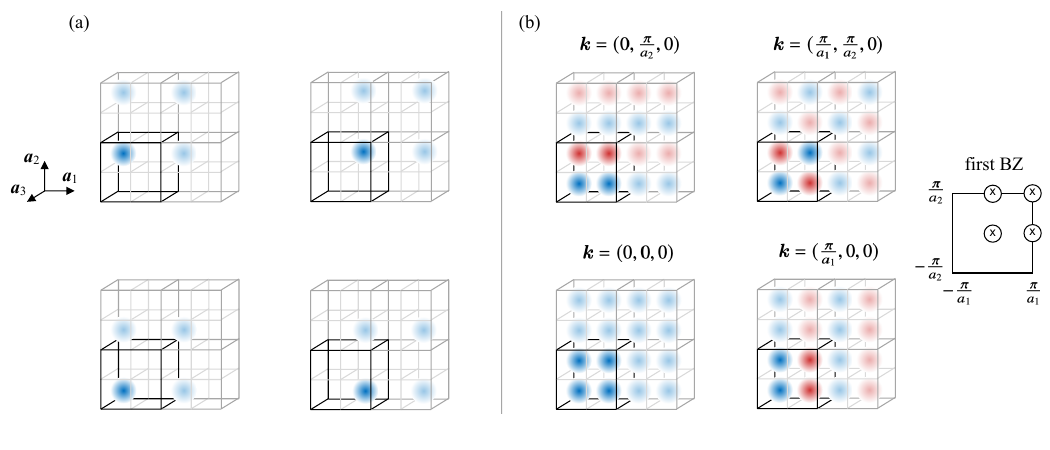}
	\caption{
Demonstration of lattice translation symmetry adaptation for a $2\times 2\times 1$ supercell (black boundaries).
In (a), we show four periodic AO basis functions, constructed from
non-periodic spherical AOs. These periodic AOs are not eigenfunctions of the unit cell lattice translation operator.
At the left, we show the unit cell lattice vectors.
In (b), we show the symmetry-adapted linear combination of the periodic AOs in (a),
which are eigenfunctions of the unit cell lattice translation operator with eigenvalues defined by the crystal momentum
$\vk$, which is indicated.
At the right, we show the four corresponding $k$-points sampled from the first Brillouin zone (BZ).
}
	\label{fig:orbitals}
\end{figure*}

\subsection{Periodic basis functions with $k$-points}
\label{ssec:kbasis}

To take advantage of lattice translation symmetry in supercell calculations,
we need basis functions that satisfy Bloch's theorem, i.e.,
\begin{equation}
\phi_{b\vk}(\vr-\vT) = e^{-i\vk\cdot\vT} \phi_{b\vk}(\vr).
\end{equation}
This can be straightforwardly accomplished with PW basis functions,
\begin{equation}
\label{eq:pw_k}
\phi_{\vG\vk}(\vr) = \frac{1}{\sqrt{N_k\Omega}} e^{i(\vk+\vG)\cdot\vr}
\end{equation}
where $\vG$ is a reciprocal lattice vector of the unit cell, 
and with periodic AO basis functions,
\begin{equation}
\label{eq:gto_k}
\phi_{\mu\vk}(\vr) = \sum_\vT e^{i\vk\cdot\vT} \tilde{\phi}_\mu(\vr-\vT),
\end{equation}
both of which can be easily checked to be eigenfunctions of the lattice
translation operator with crystal momentum $\vk$.  Comparing to the basis
functions presented in Sec.~\ref{sec:basis}, we see that periodic basis
functions of a single unit cell have $\vk=\bm{0}$, which is commonly called the
$\Gamma$ point.

We want to address a common source of confusion for newcomers to the field,
which is the relationship between PW momenta $\vG$ and crystal momenta $\vk$.
Operators that exhibit continuous (infinitesimal) translation symmetry, such as the kinetic energy
or the pairwise Coulomb or Ewald interactions, conserve the total PW momentum $\vG$.
Operators that exhibit lattice translation symmetry, such as the electron-nuclear interaction,
conserve the total crystal momentum $\vk$ (modulo $\vG$). The full Hamiltonian exhibits the lower of
the two symmetries, namely lattice translation symmetry, and therefore
conserves only crystal momentum. Nonetheless, PWs are valuable basis functions,
but their momentum $\vG$ is not conserved due to the underlying atomic lattice.

We stress that the basis functions in Eqs.~(\ref{eq:pw_k})
and (\ref{eq:gto_k}) are basis functions of the \textit{supercell} containing
$N_k$ unit cells.
They are simply a special (symmetry adapted) choice of supercell basis functions that are
eigenfunctions of the lattice translation operator. Let us explore this observation
for PWs and periodic AOs.

Returning to one dimension for notational simplicity, consider a
supercell containing two unit cells. The unit cell has lattice constant $a$ and
the supercell has lattice constant $2a$. For a PW basis with some kinetic energy cutoff,
both supercell momenta $G_1^{(\mathrm{S})}=0$ and $G_2^{(\mathrm{S})}=2\pi/(2a)$ belong to the basis.
However, these can be reinterpreted as both having $G=0$ but $k_1=0,$ and $k_2=\pi/a$, respectively.
Therefore, these two PW basis functions transform differently under a unit cell lattice translation (the first is unchanged and the second changes sign),
but this property and its computational implications were not clear in the original supercell picture.

Now consider periodic AOs for the same one-dimensional supercell. Periodic
supercell basis functions centered on atoms in either unit cell belong to the basis, i.e.,
\begin{subequations}
\begin{align}
\phi_{\mu,1}^{(\mathrm{S})}(r) &= \sum_{n} \tilde{\phi}_\mu(r-2na), \\
\phi_{\mu,2}^{(\mathrm{S})}(r) &= \sum_{n} \tilde{\phi}_\mu(r-a-2na),
\end{align}
\end{subequations}
neither of which is an eigenstate of the unit cell lattice translation operator.
However, the linear combinations,
\begin{subequations}
\begin{align}
\phi_{\mu,k_1}^{(\mathrm{S})}(r) &= \phi_{\mu,1}^{(\mathrm{S})}(r) + \phi_{\mu,2}^{(\mathrm{S})}(r)
    = \sum_{n} \tilde{\phi}_\mu(r-na), \\
\phi_{\mu,k_2}^{(\mathrm{S})}(r) &= \phi_{\mu,1}^{(\mathrm{S})}(r) - \phi_{\mu,2}^{(\mathrm{S})}(r)
    = \sum_n e^{i\pi n} \tilde{\phi}_\mu(r-na),
\end{align}
\end{subequations}
take the form of Eq.~(\ref{eq:gto_k}) and
are eigenfunctions of the lattice translation operator with $k_1 = 0$ and $k_2 = \pi/a$.
An example of this translational symmetry adaptation in a $2\times 2\times 1$ supercell
(i.e., with $2\times 2\times 1$ $k$-point sampling) is shown in Fig.~\ref{fig:orbitals}.

Therefore, because the basis functions of a supercell or a unit cell with $k$-points
are identical (for PWs) or related by a unitary transformation (for AOs),
calculations using either choice must agree exactly for basis-independent quantities
such as the total energy or the canonical orbital energies.
The symmetry-adapted set is preferable for the computational
savings that it immediately implies, as discussed more below. (However, approximate wavefunctions can
have a symmetry that is lower than that of the Hamiltonian. A familiar
example is the family of spin-unrestricted methods, which break $S^2$ symmetry. The use
of basis functions with $k$-points enforces lattice translation symmetry,
whereas the use of unadapted supercell basis functions could allow the breaking of
lattice translation symmetry, as realized in charge-density or spin-density waves.)

To summarize, it is useful to remember that every calculation with a uniform
$k$-point mesh is actually a calculation of the equivalent supercell without
$k$-points, but with lattice translation symmetry enforced.  In fact, testing
the equality of the numerical results of a calculation with $k$-points and the
equivalent supercell without $k$-points is a valuable test of implementation
correctness.

Integrals of basis functions with $k$-points can be readily evaluated using the same techniques as before.
The only modification is the guaranteed conservation of crystal momentum for all operators that are invariant to lattice translations, which includes those that appear in the Hamiltonian.
Moreover, because we are interested in observables per unit cell, and not per supercell, it is common to define integrals accordingly,
\begin{align}
\label{eq:int1e_k}
    & O_{b_1b_2}(\vk_1) = \int_\Omega d^3r\  \phi^*_{b_1\vk_1}(\vr) \hat{O} \phi_{b_2\vk_1}(\vr) \\
\begin{split}
&(b_1\vk_1, b_2\vk_2| b_3\vk_3, b_4\vk_4) \\
&\hspace{1em}= 
\int_\Omega d^3r_1 \int_{N_k\Omega} d^3r_2\ \phi^*_{b_1\vk_1}(\vr_1) \phi_{b_2\vk_2}(\vr_1) \\
&\hspace{6em}\times v_\mathrm{E}(\vr_1-\vr_2) \phi^*_{b_3\vk_3}(\vr_2) \phi_{b_4\vk_4}(\vr_2),
\end{split}
\end{align}
where the first integration is over the unit cell volume (one can easily show that integrals over the supercell volume are related to the above integrals by a factor of $N_k$).
For the ERIs, crystal momentum conservation requires $\vk_1+\vk_3 = \vk_2+\vk_4+\vG$, as already discussed, which will often be implicit in our notation.

\subsubsection{PW integrals with $k$-points}

Evaluating the above integrals for PW basis functions with $k$-points gives
\begin{align}
& T_{\vG\vG'}(\vk) = \frac{|\vk+\vG|^2}{2N_k} \delta_{\vG\vG'} \\
& V_{\vG\vG'}(\vk) = - (N_k\Omega)^{-1} v_\mathrm{E}(\vG-\vG')\rho_\mathrm{n}(\vG-\vG') \\
\begin{split}
& (\vG_1\vk_1, \vG_2\vk_2 | \vG_3\vk_3, \vG_4\vk_4) \\
&\hspace{1em}=
    (N_k^2\Omega)^{-1} v_\mathrm{E}(\vk_{34}+\vG_{34})
       \delta_{\vG_1+\vG_3,\vG_2+\vG_4},
\end{split}
\end{align}
where $\vk_{34} = \vk_4-\vk_3$.

\subsubsection{Periodic AO integrals with $k$-points}

For symmetry-adapted periodic AO basis functions, integrals can in principle be
evaluated via lattice summations of integrals of non-periodic basis functions
weighted by phase factors,
\begin{align}
&O_{\mu\nu}(\vk) = \sum_{\vT} e^{i\vk\cdot \vT} O_{\mu\nu}^{\bm{0}\vT} \\
\begin{split}
&(\mu\vk_1, \nu\vk_2 | \kappa\vk_3, \lambda\vk_4) \\
&\hspace{1em} = 
    \sum_{\vT_2\vT_3\vT_4} e^{i(\vk_2\cdot\vT_2 -\vk_3\cdot\vT_3 + \vk_4\cdot\vT_4)}
    (\mu^{\bm{0}} \nu^{\vT_2} | \kappa^{\vT_3} \lambda^{\vT_4}).
\end{split}
\end{align}
As before, it is most efficient to evaluate the long-range part of the one- and
two-electron Ewald integrals in reciprocal space. For example,
\begin{equation}
\label{eq:eris_kpts}
\begin{split}
&(\mu\vk_1, \nu\vk_2 | \kappa\vk_3, \lambda\vk_4)^\mathrm{LR} \\
    &\hspace{1em} = \frac{1}{N_k^2\Omega}
    \sum_{\vG}
    \rho_{\mu\vk_1,\nu\vk_2}(\vk_{12}-\vG) v_\mathrm{E}^\mathrm{LR}(\vk_{34}+\vG) \\
    &\hspace{6em} \times \rho_{\kappa\vk_3,\lambda\vk_4}(\vk_{34}+\vG).
\end{split}
\end{equation}

Because of crystal momentum conservation, the size of the ERI tensor in
periodic AO basis sets
is $O(N_k^3 N_\mathrm{AO}^4)$.
Although this is smaller, by a factor of $N_k$, than the size of the tensor with lattice translation
symmetry neglected, it is still computationally prohibitive for systems with moderately
sized basis sets and/or $k$-point meshes. For example, a system described by $N_\mathrm{AO} = 50$
basis functions per unit cell and $N_k = 2\times2 \times2$ $k$-points requires over 50~GB to store
the ERIs (ignoring permutation symmetries); the same system with $N_k = 3\times 3\times 3$ requires
almost 2~TB.
One solution to this storage problem is to recalculate the ERIs as they are needed
(so-called integral-direct approaches)~\cite{Sun2023}.
Another solution is to find a compressed representation of the ERI tensor, which we describe next.

\section{Periodic ERI compression}
\label{sec:compression}

\subsection{Density fitting}
\label{ssec:df}

Just like in molecular quantum chemistry, the density fitting (DF)
approximation~\cite{Kendall1997}, also known as resolution of the identity, is
extremely useful for handling the four-index ERIs in periodic calculations with
AO basis functions.  As can be seen in Eqs.~\ref{eq:pwdf} and
\ref{eq:eri_gpw}, calculations using an AO basis with GPW integration can be
understood as density fitted using an auxiliary basis of PWs.
In this section, we focus on AO calculations with an auxiliary basis of $N_\mathrm{aux}$ AOs,
which is the traditional DF approach in molecular quantum chemistry.
In the remainder of this section, all formulas will be presented assuming that
symmetry-adapted AO basis functions with $k$-points are used, and
the auxiliary AOs take the same form,
\begin{equation}
\chi_{P\vk}(\vr) = \sum_\vT e^{i\vk\cdot\vT} \tilde{\chi}_P(\vr-\bm{T} ).
\end{equation}
The product of two symmetry-adapted basis functions is an eigenstate of the lattice
translation operator with a momentum equal to the sum of the two momenta,
\begin{equation}
\phi_{\mu\vk_1}^*(\vr-\vT) \phi_{\nu\vk_2}(\vr-\vT)
    = e^{i(-\vk_1+\vk_2)\cdot\vT} \phi_{\mu\vk_1}^*(\vr) \phi_{\nu\vk_2}(\vr).
\end{equation}
The product can therefore be expanded using auxiliary basis functions with
momentum $\vk_{12} = -\vk_1+\vk_2$,
\begin{equation}
\label{eq:gdf}
\phi_{\mu\vk_1}^*(\vr) \phi_{\nu\vk_2}(\vr)
\approx \sum_{P} d_{P\vk_{12}}^{\mu\vk_1,\nu\vk_2} \chi_{P\vk_{12}}(\vr).
\end{equation}
For a uniform $k$-point mesh including
$\vk_1$ and $\vk_2$, the momentum difference $\vk_{12}$
is also in the mesh, and therefore the crystal momenta of the auxiliary basis
set are naturally chosen from the same $k$-point mesh as the primary basis set.

In DF, the fitting coefficients are most commonly determined using the
Coulomb metric, leading to
\begin{equation}
d_{P\vk_{12}}^{\mu\vk_1,\nu\vk_2} = \sum_{Q} \left[\mathbf{J}^{-1}(\vk_{12})\right]_{PQ}
    (Q,-\vk_{12}|\mu\vk_1,\nu\vk_2),
\end{equation}
and the ERI approximation
\begin{gather}
\label{eq:gdf_eri}
(\mu\vk_1,\nu\vk_2|\kappa\vk_3,\lambda\vk_4) \approx \sum_{P}
    L_{P\vk_{12}}^{\mu\vk_1,\nu\vk_2} L_{P\vk_{34}}^{\kappa\vk_3,\lambda\vk_4}, \\
L_{P\vk_{12}}^{\mu\vk_1,\nu\vk_2} = \sum_Q d_{Q\vk_{12}}^{\mu\vk_1,\nu\vk_2}
    \left[\mathbf{J}^{1/2}(\vk_{12})\right]_{PQ}
\end{gather}
(recall that $\vk_{12} = -\vk_{34} + \vG$ due to momentum conservation).
The cost of periodic DF is determined by the calculation and storage
of the two-center integrals
\begin{equation}
\label{eq:JPQ}
J_{PQ}(\vk) = \int_\Omega d^3r_1 \int_{N_k\Omega} d^3 r_2\
    \chi_{P,{-\vk}}(\vr_1) v_\mathrm{E}(r_{12}) \chi_{Q\vk}(\vr_2)
\end{equation}
and three-center integrals
\begin{equation}
\begin{split}
&(Q,-\vk_{12}|\mu \vk_1, \nu\vk_2) \\
&\hspace{0em} = \int_\Omega d^3r_1 \int_{N_k\Omega} d^3r_2\ \chi_{Q,-\vk_{12}}(\vr_1)
   v_\mathrm{E}(r_{12}) \phi_{\mu\vk_1}^*(\vr_2) \phi_{\nu\vk_2}(\vr_2).
\end{split}
\end{equation}
These integrals are most efficiently evaluated using the range-separated form
of the Ewald potential, analogous to Eq.~(\ref{eq:eri_ewald}), along with
integral screening approximations~\cite{Ye2021,Ye2021a}.  The three-center
integrals are responsible for $O(N_k^2 N_\mathrm{AO}^2 N_\mathrm{aux})$ storage
cost, which is clearly better than the $O(N_k^3 N_\mathrm{AO}^4)$ storage cost
required for the four-center ERIs, as long as
$N_\mathrm{aux} \propto N_\mathrm{AO}$.  
(periodic DF calculations can also be performed in an integral-direct fashion to avoid storage~\cite{Bintrim2022}).
Moreover, while the scaling of the
computation time is typically unchanged with DF, the prefactor is commonly
much smaller.
As we already emphasized at the end of Sec.~\ref{ssec:gto}, the evaluation of
these DF integrals is quite expensive due to lattice summations and Fourier
transforms; even though they only have to be calculated once at the beginning
of a periodic quantum chemistry calculation, they can still take a large
fraction of the total time.

\subsection{Stronger ERI compressions}
\label{ssec:stronger_compression}

Even with DF, integral storage and manipulation can be limiting for large systems,
which can be alleviated through more aggressive compression strategies, although
these typically incur a larger error.
For example, the local DF approximation can be used, which restricts the auxiliary
basis functions to be near the atomic orbitals of the pair density,
\begin{equation}
\phi_{\mu\vk_1}^*(\vr) \phi_{\nu\vk_2}(\vr) \approx \sum_{P \in [\mu\nu]}
    d_{P\vk_{12}}^{\mu\vk_1,\nu\vk_2} \chi_{P\vk_{12}}(\vr),
\end{equation}
where $[\mu\nu]$ indicates a local domain of auxiliary basis functions.
The most aggressive version of local DF uses only auxiliary basis
functions centered on the same atoms as $\mu$ and $\nu$ (known
as the pair-atom resolution of the identity or concentric atom DF).
In recent years, local DF has been applied to lower the cost of several methods in periodic quantum
chemistry~\cite{Ihrig2015,Wang2020a,Kokott24JCP,Bussy24JCP}.

A second example of stronger ERI compression is the
interpolative separable density fitting (ISDF) approach~\cite{Lu2015}, which is a flavor
of tensor hypercontraction~\cite{Hohenstein2012}. These methods have emerged
only in the last ten years and are still under development,
especially for periodic systems~\cite{Hummel2017,Rettig2023,Smyser2024}.  In the ISDF
approximation with $k$-points, the orbital product is approximated as
\begin{equation}
\label{eq:isdf}
\phi_{\mu\vk_1}^*(\vr) \phi_{\nu\vk_2}(\vr)
\approx \sum_{P} \phi_{\mu\vk_1}^*(\vr_P) \phi_{\nu\vk_2}(\vr_P) \xi_{P\vk_{12}}(\vr),
\end{equation}
where $\vr_P$ are a set of interpolation points and $\xi_{P\vk}(\vr)$ are interpolation functions
(in practice, the interpolation functions are commonly chosen to be independent of $\vk$).
Compared to the traditional DF approximation in Eq.~(\ref{eq:gdf}), the ISDF
approximation separates the orbital indices of the expansion coefficents, and the
fitting functions are determined numerically rather than being predefined.
The interpolation points are chosen automatically,
after which the interpolation functions are obtained by a least-squares fit.
Once the ISDF approximation is achieved, the ERIs take the compressed form
\begin{equation}
\label{eq:isdf_eri}
\begin{split}
&(\mu\vk_1,\nu\vk_2|\kappa\vk_3,\lambda\vk_4) \\
&\hspace{1em} \approx \sum_{PQ} \phi_{\mu\vk_1}^*(\vr_P) \phi_{\nu\vk_2}(\vr_P)
    J_{PQ}(\vk_{12}) \phi_{\kappa\vk_3}^*(\vr_Q) \phi_{\lambda\vk_4}(\vr_Q),
\end{split}
\end{equation}
where $J_{PQ}(\vk)$ has the same form as in Eq.~(\ref{eq:JPQ}).
The storage requirement
for the ISDF factorization in Eq.~(\ref{eq:isdf_eri}) scales as
$O(N_k N_\mathrm{ISDF}^2)$, which is lower than the DF approximation as long
as the number of interpolation grid points $N_\mathrm{ISDF}$ grows linearly
with system size.  However, a large number of interpolation points can make the
memory requirements high, and the computational cost to achieve the
decomposition in Eq.~(\ref{eq:isdf}), which scales cubically with system size,
can be dominant.  For more details on the theory and applications of ISDF, we
refer to a recent review article~\cite{Qin2023} and references therein.

\section{Example wavefunction theories, computational cost, and finite-size errors}
\label{sec:theory}

In this section, we present formulas for a few common periodic wavefunction theories, namely
HF, MP2, and CCSD, focusing on AO calculations, but some discussion of PW calculations
will also be provided.
We assume $N_k$ $k$-points sampled from the Brillouin zone, $N$ electrons per unit cell, 
and $N_\mathrm{AO}$ ($N_\mathrm{PW}$) basis functions per unit cell in AO (PW) calculations.
For simplicity, we assume a closed-shell system treated with a spin-restricted
formalism.  
We use the standard notation that $i,j,\ldots$ indicate the $N_\mathrm{occ} = N/2$ occupied orbitals,
$a,b,\ldots$ indicate the $N_\mathrm{vir} = N_\mathrm{AO}-N/2$ virtual (unoccupied) orbitals, and $p,q,\ldots$ indicate the $N_\mathrm{AO}$
general crystalline orbitals (all per unit cell).
With DF, we assume $N_\mathrm{aux} \propto N_\mathrm{AO}$.

We reiterate that the expressions and the associated cost savings
are essentially identical to those that exploit other symmetries, such as point
group symmetries in molecular quantum chemistry.

\subsection{Hartree-Fock}

With periodic boundary conditions and $k$-points, the HF Roothaan-Hall equation is
\begin{equation}
\mathbf{F}(\vk) \mathbf{C}(\vk) = \mathbf{S}(\vk)\mathbf{C}(\vk) \bm{\varepsilon}(\vk)
\end{equation}
where the Fock matrix is
\begin{equation}
F_{\mu\nu}(\vk) = T_{\mu\nu}(\vk) + V_{\mu\nu}(\vk) + J_{\mu\nu}(\vk) - \tfrac{1}{2}K_{\mu\nu}(\vk)
\end{equation}
and the Coulomb and exchange matrices are
\begin{align}
\begin{split}
J_{\mu\nu}(\vk) &= \frac{1}{N_k} \sum_{\vk'i} (\mu\vk, \nu\vk|i\vk',i\vk')  \\
    &= \frac{1}{N_k} \sum_{\vk'\kappa\lambda} (\mu\vk, \nu\vk | \kappa\vk', \lambda\vk') D_{\lambda\kappa}(\vk')
\end{split} \label{eq:hf_j} \\
K_{\mu\nu}(\vk) &= \frac{1}{N_k} \sum_{\vk'\kappa\lambda} (\mu\vk, \lambda\vk' | \kappa\vk', \nu\vk) D_{\lambda\kappa}(\vk') \label{eq:hf_k}
\end{align}
with the density matrix
\begin{equation}
D_{\mu\nu}(\vk) = 2\sum_{i} C_{\mu i}(\vk) C^*_{\nu i}(\vk).
\end{equation}
Note that with our convention for one-electron integrals [Eq.~(\ref{eq:int1e_k})], the crystalline orbitals are normalized to the unit cell,
\begin{equation}
    \int_\Omega d^3r\ \psi_{p\vk}^*(\vr) \psi_{q\vk}(\vr) = \delta_{pq},
\end{equation}
assuming that $\mathbf{C}(\vk)^\dagger \mathbf{S}(\vk) \mathbf{C}(\vk) = \mathbf{1}$.
The HF electronic energy per unit cell is then
\begin{equation}
\label{eq:ehf}
\begin{split}
E_0 &= \frac{N v_\mathrm{M}^{(N_k)}}{2} + \frac{1}{N_k} \sum_{\vk\mu\nu} \left[T_{\mu\nu}(\vk) + V_{\mu\nu}(\vk)\right] D_{\nu\mu}(\vk) \\
&\hspace{1em} + \frac{1}{2N_k} \sum_{\vk\mu\nu} \left[J_{\mu\nu}(\vk) - \tfrac{1}{2}K_{\mu\nu}(\vk)\right] D_{\nu\mu}(\vk),
\end{split}
\end{equation}
where $v_\mathrm{M}^{(N_k)}$ is the Madelung constant of the supercell containing $N_k$ unit cells (which in general depends on the shape, i.e., on $N_{k_x}, N_{k_y}, N_{k_z}$, but for uniform scaling is given by $v_\mathrm{M}^{(N_k)} = v_\mathrm{M}^{(1)}/N_k^{1/3}$).
Analyzed as a function of the crystal momentum $\vk$, the orbital energies $\varepsilon_p(\vk)$
define the band structure of the solid.

The computational bottleneck of periodic HF in AO basis sets is building the
Coulomb and exchange matrices at each iteration of the self-consistent field (SCF) cycle.
Both steps have a cost scaling of
$O(N_k^2 N_\mathrm{AO}^4)$ when using the four-center ERIs as in \cref{eq:hf_j,eq:hf_k}.
Using DF to approximate the ERIs as in \cref{eq:gdf_eri}, the cost scaling of
building the Coulomb matrix can be reduced to $O(N_k N_\mathrm{AO}^3)$ in a two-step
algorithm
\begin{subequations}
\begin{align}   \label{eq:hf_j_df}
    &v_P
        = \frac{1}{N_k} \sum_{\vk \kappa\lambda}
        L_{P\bm{0}}^{\kappa\vk,\lambda\vk}
        D_{\lambda\kappa}(\vk)  \\
    &J_{\mu\nu}(\vk)
        = \sum_{P}^{N_{\textrm{aux}}}
        L_{P\bm{0}}^{\mu\vk,\nu\vk} v_P
\end{align}
\end{subequations}
By contrast, using DF for building the exchange matrix
\begin{subequations}
\begin{align}   \label{eq:hf_k_df}
    &W_{P\vk_{12}}^{\mu\vk_1, i\vk_2}
        = \sum_{\lambda} L_{P\bm{0}}^{\mu\vk_1,\lambda\vk_2} C_{\lambda i}(\vk_2)   \\
    &K_{\mu\nu}(\bm{k}_1)
        = \frac{2}{N_k} \sum_{\vk_2 i P}
        W_{P\vk_{12}}^{\mu\vk_1 i\vk_2}
        W_{P\vk_{12}}^{\nu\vk_1 i\vk_2*}
\end{align}
\end{subequations}
only reduces the prefactor but leaves the $O(N_k^2 N_\mathrm{AO}^4)$ cost scaling unchanged.
Like in molecular quantum chemistry, the scaling with $N_\mathrm{AO}$ can be reduced
with additional approximations, such as integral- and density-matrix based screening
or local density fitting.

In a PW basis, the Fock matrix $F_{\vG\vG'}(\vk)$ is defined analogously, using the
integrals derived in Sec.~\ref{ssec:pw}.
However, the number of PWs used is typically much larger than the number of
AOs used for the same accuracy. The large number of PWs makes these matrices
large, and complete diagonalization at every iteration of the self-consistent
field cycle is impractical. Therefore, in practice, only the occupied orbitals
(and optionally a subset of the unoccupied orbitals) are found via iterative
eigenvalue algorithms, such as Davidson diagonalization, which require
matrix-vector products, such as
\begin{equation}
\begin{split}
[\mathbf{K}(\vk)\mathbf{C}_i(\vk_2)]_{\vG_2} &= \frac{1}{\Omega N_k} \sum_{j\vG_1\vk_1} \rho_{j\vk_1,i\vk_2}(\vk_{12}+\vG_{12}) \\
    &\hspace{2em}\times v_\mathrm{E}(\vk_{12}+\vG_{12}) C_{\vG_1 j}(\vk_1),
\end{split}
\end{equation}
As written, the above expression (including the Fourier transform of the orbital pairs) scales like $O(N_\mathrm{occ}^2 N_\mathrm{PW}^2 N_k^2)$, although the scaling
with respect to $N_\mathrm{PW}$ can be reduced to $O(N_\mathrm{PW} \ln N_\mathrm{PW})$ by using several fast Fourier
transforms with real-space quadrature.
The cost of this exchange matrix-vector product, which dominates the scaling of HF and hybrid DFT calculations,
is due to the nonlocality of the exchange operator
and can be reduced using the adaptively compressed exchange (ACE) method~\cite{Lin2016}.
See Ref.~\onlinecite{Wu2021} for details of an efficient combination of ACE and ISDF
for PW-based HF or hybrid DFT calculations.

\begin{figure}[b]
	\includegraphics[scale=1.0]{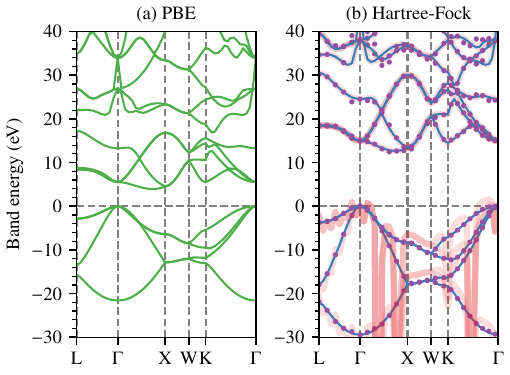}
	\caption{
Band structure of diamond with a $5\times 5\times 5$ $k$-point mesh
using PBE (a) and HF (b).  The PBE band structure was
calculated non-self-consistently along the path following a single self-consistent calculation. The HF band structure in
faded red was calculated the same way using the Madelung-corrected Ewald
potential, demonstrating the discontinuities. The HF band structure was
calculated more accurately in two ways: self-consistently with the $k$-point mesh shifted to include each point
along the band path (blue lines), and using a Wannier interpolation of the Fock
matrix (purple circles).  Calculations use a GTH
pseudopotential~\cite{Goedecker96PRB,Hartwigsen98PRB} and the GTH-cc-pVDZ basis
set~\cite{Ye2022}.
}
\label{fig:diamond_bands}
\end{figure}

When calculating the band structure $\varepsilon_p(\vk)$, one commonly wants a
dense set of $k$-points along a path through special high-symmetry points in
the Brillouin zone. For local DFT calculations, these eigenvalues can be
calculated non-self-consistently with an electron density calculated
self-consistently on a coarse $k$-point mesh. For example,
the Coulomb matrix can be calculated at a $k$-point $\vk^*$ that does not belong to the mesh,
\begin{equation}
J_{\mu\nu}(\vk^*) = \frac{1}{N_k} \sum_{\vk \kappa\lambda} (\mu\vk^*,\nu\vk^*|\kappa\vk,\lambda\vk) D_{\lambda\kappa}(\vk)
\end{equation}
using the self-consistent density matrix $D_{\lambda\kappa}(\vk)$.
The needed ERI is given by Eq.~(\ref{eq:eris_kpts}) with $\vk_1=\vk_2=\vk^*$ and
$\vk_3=\vk_4=\vk$, which requires evaluating the Ewald potential only as
$v_\mathrm{E}(\vG)$.

However, when evaluating the exchange matrix $K_{\mu\nu}(\vk^*)$, the ERIs
require us to evaluate the Ewald potential as $v_\mathrm{E}(\vk^*-\vk+\vG)$,
which highlights the discontinuity of the Ewald potential in reciprocal space
at $\vG=\bm{0}$ and generates discontinuities in the band structure.
Specifically, when $\vk^* = \vk$, we have $v_\mathrm{E}(\bm{0}) = 0$, but when
$\vk^* = \vk + \bm{\delta}$, we have $v_\mathrm{E}(\bm{\delta}) = 4\pi/\delta^2$,
which is very large.
This discontinuity only affects the occupied band energies, because the
unoccupied exchange matrix element $K_{aa}(\vk^*)$ has a suppressed $\vG=0$ contribution
due to the orthogonality of the occupied and unoccupied orbitals.

To generate a smooth band structure with nonlocal exchange, one has several
options: (1) shift the entire $k$-point mesh to include the band point $\vk^*$
and perform SCF calculations along the entire path; (2) use a representation of the Coulomb
potential that is continuous in reciprocal space at $\vG=\bm{0}$, such as real-space
truncations~\cite{Spencer2008,Sundararaman2013}; (3) perform Wannier
interpolation~\cite{Yates07PRB,Mostofi08CPC} of the Fock matrix.
To perform Wannier interpolation after a
$k$-point sampled SCF calculation on a coarse mesh, one generates Wannier functions associated
with the unit cell at $\vT$ in the $k$-point sampled supercell,
\begin{equation}
\label{eq:wannier}
w_{\tilde{\mu}\vT}(\vr)
    = \frac{1}{N_k} \sum_{p\vk} e^{-i\vk\cdot\vT} U_{p\tilde{\mu}}(\vk) \psi_{p\vk}(\vr).
\end{equation}
The freedom in the unitary matrix $U_{p\tilde{\mu}}(\vk)$ is used to
spatially localize the orbitals by minimizing a cost function,
such as the Boys cost function (leading to so-called maximally localized
Wannier functions) or the Pipek-Mezey cost function.
The Wannier-interpolated Fock matrix at arbitrary $\vk^*$ can then be constructed
from the SCF band energies,
\begin{subequations}
\begin{align}
F_{\tilde{\mu}\tilde{\nu}}(\vk^*) &= \sum_\vT^{N_k} e^{i\vk^*\cdot\vT} F_{\tilde{\mu}\tilde{\nu}}(\vT)
\label{subeq:Fk_from_FT}
\\
F_{\tilde{\mu}\tilde{\nu}}(\vT) &= \frac{1}{N_k} \sum_{p\vk}
    e^{-i\vk\cdot\vT} U^*_{p\tilde{\mu}}(\vk) U_{p\tilde{\nu}}(\vk) \varepsilon_p(\vk),
\end{align}
\end{subequations}
and simply diagonalized to obtain an estimate of the band structure.
Note that in Eq.~(\ref{subeq:Fk_from_FT}), the $\bm{T}$ summation
can be done over the $N_k$ values in the supercell or over the $N_k$ values of
the Wigner-Seitz supercell, the latter of which shows better performance for
coarse $k$-point meshes~\cite{Yates07PRB,Mostofi08CPC} (the results
differ from one another because, although 
$F_{\tilde{\mu}\tilde{\nu}}(\vT-\vT_\mathrm{sup}) = F_{\tilde{\mu}\tilde{\nu}}(\vT)$
where $\vT_\mathrm{sup}$ is a supercell translation vector, we have 
$e^{i\vk^*\cdot\vT_\mathrm{sup}} \neq 1$ when
$\vk^*$ does not belong to the mesh).
The PBE and HF band structures of diamond are shown in Fig.~\ref{fig:diamond_bands},
where the HF band structure was calculated non-self-consistently with the Ewald potential,
self-consistently along the entire path [option (1) in the list above], and using Pipek-Mezey Wannier
interpolation~\cite{Yang2026} [option (3)].
The comparison between PBE and HF shows that the HF band gap and band width are larger, which
is a typical result.

\subsection{MP2 and coupled-cluster theory}

Both MP perturbation theory and CC theory are defined by the amplitudes of
excitation operators, $t_{i\vk_1}^{a\vk_2}$, $t_{i\vk_i, j\vk_j}^{a\vk_a b\vk_b}$, etc.,
and these excitation operators must conserve the total crystal
momentum of the reference determinant on which they act.
This reference determinant is commonly the HF one, but need not be.

Using a MP-type partitioning of the Hamiltonian,
the MP2 correlation energy is
\begin{equation}
\label{eq:emp2}
\begin{split}
E_\mathrm{c}^{(2)} &= \frac{2}{N_k} \sum_{\vk} \sum_{ia}
        t_{i\vk}^{a\vk} F_{ia}(\vk) \\
&\hspace{-2em}
+\frac{1}{N_k^3}\ \sideset{}{'}\sum_{\vk_i\vk_j\vk_a\vk_b} \sum_{ijab}
        t_{i\vk_i, j\vk_j}^{a\vk_a, b\vk_b} \\
&\hspace{0em} \times \left[ 2(i\vk_i, a\vk_a | j\vk_j, b\vk_b)
        - (i\vk_i, b\vk_b | j\vk_j, a\vk_a) \right]
\end{split}
\end{equation}
where the primed summation enforces crystal momentum
conservation, $\vk_{i}+\vk_{j} = \vk_{a}+\vk_{b}+\vG$,
and the first-order perturbed wavefunction amplitudes are
\begin{subequations}
\begin{align}
    t_{i\vk}^{a\vk}
        &= \frac{F^*_{ia}(\vk)}{\varepsilon_{i}(\vk) - \varepsilon_{a}(\vk)} \\
    t_{i\vk_i, j\vk_j}^{a\vk_a, b\vk_b}
        &= \frac{(i\vk_i, a\vk_a | j\vk_j, b\vk_b)^*}
        {
            \varepsilon_{i}(\vk_i)
            - \varepsilon_{a}(\vk_a)
            + \varepsilon_{j}(\vk_j)
            - \varepsilon_{b}(\vk_b)
        }
\end{align}
\end{subequations}
The cost of evaluating the MP2 correlation energy (\ref{eq:emp2}) scales as $O(N_k^3 N_\mathrm{occ}^2 N_\mathrm{vir}^2)$,
which is identical in either AO or PW calculations.
However, for AO calculations, the cost of transforming the ERIs from the AO basis to the crystalline orbital basis
\begin{equation}
\begin{split}
    (i\vk_i, a\vk_a | j\vk_j, b\vk_b)
        &= \sum_{\mu\nu\kappa\lambda}
        (\mu\vk_i, \nu\vk_a | \kappa\vk_j, \lambda\vk_b)  \\
        &\hspace{-1em}\times
        C_{\mu i}^*(\vk_i) C_{\nu a}(\vk_a)
        C_{\kappa j}^*(\vk_j) C_{\lambda b}(\vk_b)
\end{split}
\end{equation}
scales as $O(N_k^3 N_\mathrm{occ} N_\mathrm{AO}^4)$, which dominates the computational cost of MP2, as usual.
Using DF for the integral transform
\begin{gather}
    L_{P\vk_{ia}}^{i\vk_i a\vk_a}
        = \sum_{\mu\nu} L_{P\vk_{ia}}^{\mu\vk_i \nu\vk_a}
        C^*_{\mu i}(\vk_i) C_{\nu a}(\vk_a) \\
    (i\vk_i, a\vk_a | j\vk_j, b\vk_b)
        = \sum_{P} L_{P\vk_{ia}}^{i\vk_i, a\vk_a}
        L_{P\vk_{jb}}^{j\vk_j, b\vk_b}
\end{gather}
has a cost of $O(N_k^3 N_\mathrm{occ}^2 N_\mathrm{vir}^2 N_\mathrm{aux})$, which still has $O(N^5)$ scaling
with unit cell size, but a significantly lower prefactor, as typical of DF MP2.

The CC correlation energy has the same form as \cref{eq:emp2},
but the amplitudes are obtained by solving a system of nonlinear equations,
which determines the overall cost. For example, in CCSD, the cost is determined
by contractions such as the particle-particle ladder term,
\begin{equation}
\label{eq:ccsd}
\sideset{}{'} \sum_{\vk_c\vk_d} \sum_{cd} (a\vk_a,b\vk_b|c\vk_c,d\vk_d) t_{i\vk_i, j\vk_j}^{c\vk_c,d\vk_d},
\end{equation}
which has $O(N_k^4 N_\mathrm{occ}^2 N_\mathrm{vir}^4)$ scaling due to
crystal momentum conservation in the ERIs and the amplitudes, as indicated
by the primed summation.

\subsection{Finite-size errors}
\label{ssec:finite}

\begin{figure}
	\includegraphics[scale=1.0]{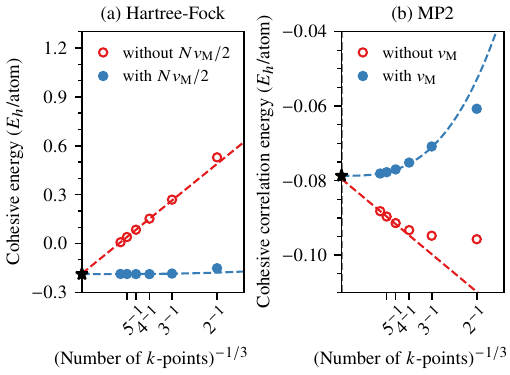}
	\caption{
Cohesive energy of diamond as a function of the number of $k$-points sampled,
up to $N_k=7\times 7\times 7$.  Calculations use a GTH
pseudopotential~\cite{Goedecker96PRB,Hartwigsen98PRB} and the GTH-cc-pVDZ basis
set~\cite{Ye2022}.  The HF cohesive energy (a) is shown without and with the
$Nv_\mathrm{M}/2$ term appearing in the total energy in \cref{eq:Vee,eq:ehf}.
The MP2 correlation energy contribution to the cohesive energy (b) is shown
without and with a shift of $v_\mathrm{M}$ to the occupied orbital energies, as
realized by the probe-charge Ewald auxiliary function treatment of the exchange
matrix.  Dashed lines are fits assuming finite-size errors that scale as
$N_k^{-1/3}$ (red) and $N_k^{-1}$ (blue).
}
\label{fig:diamond}
\end{figure}

As discussed in several times above, all periodic quantum chemistry calculations
contain a finite-size error, due to the limited size of the supercell or
$k$-point mesh. Although DFT calculations can often be performed with
thousands of $k$-points, HF and post-HF calculations have so far been restricted
to hundreds of $k$-points or less.
Although some observables
can be accurately evaluated from such calculations, it is common to attempt extrapolation
to the thermodynamic limit (i.e., $N_k\rightarrow \infty$).
Such extrapolation requires knowledge of the asymptotic form of the decay of the
finite-size error, which is commonly polynomial at these levels of theory,
\begin{equation}
E(N_k) = E(N_k\rightarrow\infty) + a N_k^{-p} + \ldots
\end{equation}
In this section, we discuss the leading-order finite-size errors, for which
$p = 1/3$ or $p=1$.
However, there is no guarantee that calculations performed with small,
accessible values of $N_k$ are in this asymptotic regime, and extrapolated
results have to be judged with caution~\cite{Moerman2025a,Moerman2025,Li2026}.  Below, we provide some introductory
discussion of finite-size errors in HF theory and correlated theories, such as
MP2 and CCSD.

To get a handle on finite-size errors, let us temporarily simplify the notation
by suppressing $k$-points and imagine performing calculations on increasingly large
supercells. In terms of the crystalline spin-orbitals $\psi_i(\vx)$ occupied in the HF determinantal
wavefunction, the two-electron contribution to the total energy is
\begin{subequations}
\begin{align}
E_2 &= E_J' + E_K' \\
E_J' &= Nv_\mathrm{M}/2 + \frac{1}{2} \sum_{i\neq j} (ii|jj) \\
E_K' &= - \frac{1}{2} \sum_{i\neq j} (ij|ji),
\end{align}
\end{subequations}
where the $Nv_\mathrm{M}/2$ term has been included in the Coulomb energy $E_J'$ due to
its origin in classical electrostatics
(here, temporarily, $i$ and $j$ are spin-orbitals).
The usual procedure is to relax the $i\neq j$ constraint,
which is allowed because the self-interactions in the Coulomb and exchange terms cancel one another.
However, let's consider these self-interactions more closely for periodic systems,
\begin{equation}
(ii|ii) = \iint_\Omega d^3r_1 d^3r_2 \rho_i(\vr_1) v_\mathrm{E}(\vr_1-\vr_2)\rho_i(\vr_2).
\end{equation}
This is the self-interaction of a periodic charge density in a
neutralizing background, which will exhibit finite-size errors due to the use of
a finite cell and the Ewald potential.
After relaxing $i\neq j$, we can add a correction to both the Coulomb and exchange energies (canceling one another),
which eliminates the leading-order finite-size error based
on classical electrostatics. Specifically, we add the difference between the
self-interaction of an isolated point charge
and that of a point charge interacting
with its periodic images and a neutralizing background,
\begin{equation}
\label{eq:deltaE}
\Delta E = \lim_{\vr\rightarrow \bm{0}} \left[ \frac{1}{r} - v_\mathrm{E}(\vr) \right]
    = -v_\mathrm{M}.
\end{equation}
This leads to an equivalent rewriting of the two-electron energy,
\begin{subequations}
\label{eq:ejek}
\begin{align}
E_2 &= E_J + E_K \\
E_J &= Nv_\mathrm{M}/2 + \frac{1}{2} \sum_{ij}\left[ (ii|jj) - v_\mathrm{M}\delta_{ij}\right]
    = \frac{1}{2} \sum_{ij}(ii|jj) \\
E_K &= - \frac{1}{2} \sum_{ij} (ij|ji) + Nv_\mathrm{M}/2.
\end{align}
\end{subequations}
In Eq.~(\ref{eq:ejek}), the constant $Nv_\mathrm{M}/2$---which originated from
classical electrostatics in the derivation of the first-quantized
Hamiltonian---is now associated with the non-classical exchange energy.  This
behavior is purely due to the convention of including self-interaction terms in
the definition of the Coulomb and exchange energies.
We emphasize that the total energy $E_2$ is identical in either formulation.

The inclusion of finite-size corrections in the exchange energy, leading to
constant corrections like the $Nv_\mathrm{M}/2$ term above, has a long history.
Restoring our notation to use $k$-points with $i,j$ occupied spatial orbitals, 
the evaluation of the exchange energy
using the Coulomb potential (\textit{not} the Ewald potential) exhibits an integrable divergence,
\begin{equation}
\begin{split}
E_K &= -\frac{1}{N_k^2} \sum_{\vk_1\vk_2} \sum_{ij} (i\vk_1,j\vk_2|j\vk_2, i\vk_1) \\
    &= -\frac{1}{\Omega N_k^2} \sum_{\vk\vq\vG} \sum_{ij} \frac{4\pi |\rho_{i\vk,j\vk+\vq}(\vq+\vG)|^2}{|\vq+\vG|^2},
\end{split}
\end{equation}
where in the second line we defined $\vk = \vk_1$ and $\vq = -\vk_1+\vk_2$.
In the limit of dense $k$-point sampling (i.e., large supercells), the sums over $\vk$ and $\vq$ become
integrals; although the integrand diverges as $1/q^2$ near $\vq = \bm{0}$ due to the
summand with $i=j$ and $\vG=\bm{0}$, the integral converges as $N_k\rightarrow
\infty$ because
\begin{equation}
    \lim_{N_k \to \infty} \frac{1}{N_k}\sum_{\vq} \frac{1}{q^2}
        = \frac{\Omega}{(2\pi)^3}\int_{\textrm{BZ}} d^3 q \frac{1}{q^2}
\end{equation}
is finite.  Simply neglecting the divergent term (as prescribed by the Ewald
potential) causes slow convergence with errors that scale as $O(N_k^{-1/3})$,
and auxiliary function methods were introduced in the late 1980s to expedite
this convergence~\cite{Gygi1986,Massidda1993}.  To the term in the integrand
with $i=j$, one adds and subtracts a function $f(\vq)$ that has the same
$1/q^2$ divergence but can be integrated analytically. One of the terms is
integrated analytically, and the other is integrated approximately on the grid,
canceling the divergence of the Coulomb potential, giving
\begin{equation}
\label{eq:exx_aux}
E_K = -\frac{1}{\Omega N_k^2} \sum_{\vk\vq\vG} \sum_{ij}
|\rho_{i\vk j\vk+\vq}(\vq+\vG)|^2 v_f(\vq+\vG)
\end{equation}
where
\begin{equation}
\label{eq:vg_ewald}
v_f(\vq+\vG) = \begin{cases}
4\pi/ |\vq+\vG|^2 & \vq+\vG \neq \bm{0} \\
-N_k\Omega C_f & \vq+\vG = \bm{0}
\end{cases}
\end{equation}
and
\begin{equation}
C_f = \frac{1}{N_k\Omega} {\sum_{\vq\vG}}' f(\vq + \vG) - \int \frac{d^3Q}{(2\pi)^3}~f(\vQ)
\end{equation}
where the primed summation indicates $\vq + \vG \neq \bm{0}$.
If $f(\vQ)$ is chosen to be
\begin{equation}
f(\vQ) = 4\pi \frac{e^{-Q^2/4\alpha^2}}{Q^2},
\end{equation}
then
\begin{equation}
\label{eq:Cf}
C_f =
    \frac{4\pi}{N_k\Omega} {\sum_{\vq\vG}}'
    \frac{e^{-|\vq+\vG|^2/4\alpha^2}}{|\vq+\vG|^2}
    - 2\sqrt{\frac{\alpha^2}{\pi}}.
\end{equation}
With this choice, $C_f$ is the potential at the origin due to the periodic
images of a unit Gaussian charge density and its uniform neutralizing density.

Calculating exchange matrices and energies using \cref{eq:vg_ewald,eq:Cf} is
sometimes known as the probe-charge Ewald
method~\cite{Paier2005,Broqvist2009,Sundararaman2013}.  In principle, any value
of $\alpha$ can be used in the probe-charge Ewald method. When $\alpha
\rightarrow \infty$ such that the charge density is that of a point charge,
then $C_f = v_\mathrm{M}^{(N_k)}$, the Madelung constant of the supercell, and
the exchange energy calculated with Eq.~(\ref{eq:exx_aux}) is identical to that
with Eq.~(\ref{eq:ejek}).
This method is the current default
behavior in VASP (with $\alpha < \infty$) and PySCF (with $\alpha = \infty$)
for Hartree-Fock and hybrid DFT calculations.
(To control the treatment of this exact exchange divergence in periodic HF
calculations, PySCF uses an optional keyword argument \texttt{exxdiv}. Perhaps
confusingly, calculations with \texttt{exxdiv=None} use the uncorrected Ewald
potential, and calculations with \texttt{exxdiv='ewald'} (the default) apply a
Madelung constant correction, i.e., the probe-charge Ewald method with $\alpha=\infty$.)

The probe-charge Ewald method, including the simple expression in
Eq.~(\ref{eq:ejek}), produces an exchange energy (and a total energy) whose
finite-size errors decay asymptotically as $N_k^{-1}$.  In
Fig.~\ref{fig:diamond}(a), we show the convergence of the HF cohesive energy of
diamond as a function of the number of $k$-points sampled using PySCF.  Clearly, the
$Nv_\mathrm{M}/2$ term makes an enormous contribution to the energy and should
not be neglected in calculations with exact, nonlocal exchange.

Although these two approaches give the same energies, they yield
different exchange matrices and therefore different Fock matrices
$F_{\mu\nu}(\vk)$ and orbital energies $\varepsilon_p(\vk)$. In the
probe-charge Ewald approach, the occupied
orbital energies are uniformly shifted by $C_f$ (which is a shift to
lower energy), and the unoccupied orbital energies are unchanged, such
that the band gap is increased by $|C_f|$. The occupied
orbital energies converge fastest to their thermodynamic limit when the
shift is included (as $N_k^{-1}$ with this shift and as
$N_k^{-1/3}$ without this shift).
This behavior can also be justified by recalling that the orbital energies
approximate the energies associated with electron addition or removal.

Because the two approaches yield different Fock matrices, they generate
different correlation energies when used with finite-order M{\o}ller-Plesset perturbation theory,
which uses the Hamiltonian partitioning $H = H_0 + H_1$ with $H_0 = F + \langle H - F\rangle$.
The energies will agree in the thermodynamic limit, but their finite-size errors
scale differently. Specifically,
the MP2 correlation energy converges as $N_k^{-1}$ with the occupied orbital shift,
as opposed to $N_k^{-1/3}$ without the shift.
In Fig.~\ref{fig:diamond}(b), we show results for the MP2 correlation energy of
diamond calculated without and with this shift to the orbital energies
appearing in the denominator, confirming that the latter converges faster
asymptotically.
Finally, we note that several other popular methods exist for expediting the
convergence of the exchange energy to the thermodynamic limit, such as
the use of a truncated Coulomb interaction~\cite{Spencer2008,Sundararaman2013},
which may simultaneously lower the cost of calculations.

As discussed in Sec.~\ref{sec:ham},
this ambiguity makes it difficult to compare results away from the thermodynamic limit, i.e.,
with a finite $k$-point mesh or supercell size.
One possible solution is to reorganize the terms of the potential energy derived in Sec.~\ref{sec:ham}
as follows~\cite{Fraser1996},
\begin{subequations}
\label{eq:v_shifted}
\begin{align}
\bar{E}_\mathrm{nn} &= \frac{1}{2}\sum_{A\neq B} Z_A Z_B \bar{v}_\mathrm{E}(\vR_A-\vR_B) \\
\bar{V}_\mathrm{en} &= -\sum_{nA} Z_A \bar{v}_\mathrm{E}(\vr_n-\vR_A) \\
\bar{V}_\mathrm{ee} &= \frac{1}{2} \sum_{n\neq n'} \bar{v}_\mathrm{E}(\vr_n-\vr_{n'})
\end{align}
\end{subequations}
where the modified two-body Coulomb interaction is a shifted Ewald interaction,
\begin{equation}
\bar{v}_\mathrm{E}(\vr) = v_\mathrm{E}(\vr) - v_\mathrm{M},
\end{equation}
and we have assumed a charge-neutral cell (note again that $v_\mathrm{M}$ may be the Madelung
constant of a supercell, which is different than that of the unit cell).
This regrouping of the potential energy exhibits terms with the more familiar
pairwise interactions only, i.e., without constants or one-body energies due to interactions with
periodic images.  This shifted Ewald potential is no longer zero on average,
but its short-range behavior is a closer match to the original Coulomb
potential, as seen from the limit
\begin{equation}
\lim_{\vr\rightarrow \bm{0}} \bar{v}_\mathrm{E}(\vr) = \frac{1}{r},
\end{equation}
which is true by the definition of the Madelung constant [Eq.~(\ref{eq:madelung})].
Moreover, the Fourier representation of the shifted Coulomb interaction,
\begin{equation}
\bar{v}_\mathrm{E}(\vq+\vG) = \begin{cases}
4\pi/|\vq+\vG|^2 & \vq+\vG \neq \bm{0} \\
-N_k \Omega v_\mathrm{M} & \vq+\vG = \bm{0},
\end{cases}
\end{equation}
is clearly the same as the one obtained with the auxiliary function
(probe-charge Ewald) method above.
From a software implementation perspective, the use of a shifted Ewald potential
is trivial; for example, the ERI tensor is modified as
\begin{equation}
(\mu\nu|\bar{v}_\mathrm{E}|\kappa\lambda) = (\mu\nu|v_\mathrm{E}|\kappa\lambda)
    - v_\mathrm{M} S_{\mu\nu} S_{\kappa\lambda}.
\end{equation}
Alternatively, with DF, it modifies the two-center and three-center integral tensors
in the same way, while preserving the factorized structure of the ERI compression.

For non-perturbative theories---including truncated configuration
interaction, CC theory, and AFQMC---the Hamiltonian is not partitioned and
therefore the Fock matrix and orbitals energies play no substantive role.
Therefore, the energy obtained using either
$v_\mathrm{E}(\vr)$ or $\bar{v}_\mathrm{E}(\vr)$ is identical.
To see this explicitly in the working equations for CCSD,
note that the contraction in Eq.~(\ref{eq:ccsd}), with the shifted integrals, 
would produce an additional diagonal contribution of $v_\mathrm{M} t_{i\vk_i,k\vk_j}^{a\vk_a,b\vk_b}$.
Collecting all such terms from the other contractions will show that they cancel the Madelung constant contributions
to the occupied orbital energies (diagonal elements of the Fock matrix).

Another replacement for the Ewald potential is the so-called model periodic Coulomb interaction,
which has shown good performance in diffusion Monte Carlo calculations~\cite{Williamson1997,Kent1999}.

\section{Local correlation and quantum embedding}
\label{sec:embed}

As we have discussed throughout, periodic quantum chemistry calculations can be
computationally expensive if targeting the thermodynamic limit. The use of
lattice translational symmetry via $k$-point conservation is valuable,
but typically provides computational savings of only $O(N_k)$ or $O(N_k^2)$.
Therefore, additional approximation techniques that can facilitate convergence
to the thermodynamic limit are desirable.

One such class of methods are local approximations, including integral screening,
local domains, and local DF, some of which we have briefly mentioned above.
An extension of these methods, especially for post-HF calculations,
are local correlation approximations, which exploit the fact that electron
correlation is a short-range effect.
The use of local correlation approximations has been remarkably successful
in molecular quantum chemistry, and one can expect the same in
periodic quantum chemistry (especially for insulating solids).
Roughly speaking, local correlation methods partition the total system into
fragments and either perform multiple independent correlated calculations for each
one (such as in cluster-in-molecule methods~\cite{Li06JCP,Rolik13JCP})
or perform a single correlated
calculation that approximates interfragment correlation (such as in methods
based on projected atomic orbitals~\cite{Werner11JCP} or pair natural
orbitals~\cite{Riplinger13JCP}).

Essentially all local correlation methods use localized orbitals
obtained via the Wannier transformation in Eq.~(\ref{eq:wannier}).
Wannier orbitals are still periodic with respect to supercell translation vectors, but
they are no longer eigenfunctions of the unit cell lattice translation operator; instead,
they are related to one another by lattice translations,
\begin{equation}
\begin{split}
w_{\tilde{\mu}\bm{0}}(\vr-\vT) &= \sum_{p\vk}  U_{p\tilde{\mu}}(\vk) \psi_{p\vk}(\vr-\vT) \\
    &= \sum_{i\vk} e^{-i\vk\cdot\vT}  U_{p\tilde{\mu}}(\vk) \psi_{p\vk}(\vr)
    \equiv w_{\tilde{\mu}\vT}(\vr)
\end{split}
\end{equation}
Such local correlation methods are especially appropriate for periodic systems
because the number of unique fragments is $O(1)$ due to this lattice translational
symmetry, independent of the number of $k$-points sampled from the Brillouin
zone.  In this direction, we highlight the development of periodic local
MP2~\cite{Pisani08JCC,Usvyat11JCP,Usvyat15JCP,Nejad2025,Zhu2025},
RPA~\cite{Ren12NJP}, and CC
theory~\cite{Wang19JCTC,Wang22JCTC,Li23ACR,Ye24arxiv,Ye24FD,Ye24JCTC}.

The local correlation philosophy is very closely related to that of quantum
embedding methods, such as density embedding~\cite{Manby12JCTC,Yu17Book},
density matrix embedding theory~\cite{Wouters16JCTC}, and dynamical mean-field
theory~\cite{Georges96RMP}, the latter two of which have been recently
implemented in a so-called full-cell formalism~\cite{Cui20JCTC,Zhu20JCTC,Zhu21PRX} that is most
in line with ab initio quantum chemistry.
This framework has enabled applications to strongly correlated problems in materials science,
including the cuprates~\cite{Cui22Science,Cui25NC} and magnetic impurities in
metals~\cite{Zhu2024}.
In both local correlation and quantum embedding
methods, convergence to the exact result (for a given level of theory) is
achieved by increasing the size of the fragment.
The biggest qualitative difference from local correlation
methods is that quantum embedding methods typically enforce self-consistency:
commonly a one-electron description of the total system is constructed to be
self-consistent with a correlated description of a local fragment, allowing
locally correlated electronic structure to improve the
global mean-field description.  Perhaps more importantly, the nonlinear behavior achievable
through self-consistency is necessary for predicting phase transitions in extended materials.

\section{Conclusions}
\label{sec:conc}

We have given an overview of the theory and implementation
underlying wavefunction-based periodic quantum chemistry, with an emphasis on
its differences from molecular quantum chemistry.  We have focused only on
ground-state energies, but the same formalism can be applied to calculate excitation
energies of solids. For example, equation-of-motion coupled-cluster theory has
been used to calculate band structures and band
gaps~\cite{McClain2017,Gao2020,Vo2024,Moerman2025a,Moerman2025}, as well as neutral
excitation energies and spectra~\cite{Wang2020,Gallo2021,Wang2021}.

Most applications of wavefunction-based methods to date have focused on
insulating and semiconducting materials with relatively weak, short-ranged
electron correlation.
In contrast, there have been far fewer applications to atomistic
metals~\cite{Mihm2021,Neufeld2022,Carbone24RD,Ye24JCTC} (despite
a relatively large body of work on CC theory for the uniform electron
gas~\cite{Freeman1977,Shepherd2013,Shepherd2014}), for
several physical and numerical reasons. For example, HF theory is a poor
starting point for metals, the number of $k$-points necessary to converge the
energy is much higher than for insulators, and occupied orbitals cannot be
localized as tightly. Moreover, the gapless nature of metals typically requires
fractional occupations~\cite{Gironcoli1995,Santos2023}, motivating finite-temperature many-electron wavefunction theories~\cite{Santra2017,White2018,Hummel2018} and causes an infrared divergence in the correlation
energy predicted by finite-order perturbation theories, including MP2 and
CCSD(T)~\cite{Shepherd2013,Masios23PRL,Neufeld23PRL}.

Despite recent advances, most applications of the wavefunction methods
discussed in this article have been to relatively simple solids and surfaces.
Such calculations can provide benchmark values against which more affordable
approximations can be
tested~\cite{Tsatsoulis2017,Gruber2018,Shi2023,Ye24arxiv,Ye24FD,Weiske2025}. But to enable
routine applications of ab initio methods to complex, condensed-phase systems,
additional algorithmic and software developments are clearly necessary. A
complementary, promising direction is the training of machine-learning
potentials~\cite{Behler2021} with a limited number of high-level, periodic
calculations---for example, via fine-tuning or transfer
learning~\cite{Chen2023}.  To the extent that the machine-learning potential
accurately extrapolates beyond the training data, this approach would allow
large-scale molecular dynamics and Monte Carlo simulations with energies and
forces of unprecedented accuracy.  With these ongoing and future advances,
wavefunction-based periodic quantum chemistry is poised to broaden its reach
and impact across a wider range of material classes and scientific questions.

\section*{Acknowledgements}

We thank Garnet Chan, Lin Lin, Hendrik Monkhorst, and Sandeep Sharma for discussions related to some of the theory presented here, and Arman Nejad and David Tew for comments on the manuscript.
This work was supported by the National Science Foundation under Grant No.~CHE-1848369 and by the Columbia Center for Computational Electrochemistry.
The Flatiron Institute is a division of the Simons Foundation.

\section*{Data availability statement}
The data that support the findings of this study are available from the
corresponding author upon reasonable request.

\appendix

\section{Derivation of Ewald summation formula}
\label{app:ewald}

For generality, consider a periodic system of charges $q_i$ at positions $\vr_i$, with a neutral unit cell ($\sum_i q_i = 0$).
Using the splitting in Eq.~(\ref{eq:vr_split}), the short-range, real-space energy is simply
\begin{equation}
    V_\text{SR} = \frac{1}{2} \sum_{ij} \sum_{\vT}' q_i q_j \frac{\mathrm{erfc}(\eta |\vr_{ij}-\vT|)}{|\vr_{ij}-\vT|}.
\end{equation}
The long-range energy can be evaluated in reciprocal space using the Fourier transform of $\mathrm{erf}(\eta r)/r$,
\begin{equation}
\begin{split}
    V_\text{LR} &= \frac{1}{2\Omega} \sum_{\vG\neq 0} \rho(-\vG) \frac{4\pi}{G^2}e^{-G^2/4\eta^2} \rho(\vG) \\
    &= \frac{4\pi}{2\Omega} \sum_{ij} q_i q_j \sum_{\vG\neq \bm{0}}e^{i\vG\cdot\vr_{ij}} \frac{e^{-G^2/4\eta^2}}{G^2},
\end{split}
\end{equation}
although this incorrectly includes the self-interactions, which must be subtracted
\begin{equation}
    V_\text{self} = -\lim_{r\rightarrow 0} \sum_i q_i^2 \frac{\mathrm{erf}(\eta r)}{r} =  -\frac{2 \eta}{\sqrt{\pi}} \sum_i q_i^2.
\end{equation}
Separating the $i=j$ terms of $V_\text{SR}$ and $V_\text{LR}$, the total energy can be rewritten 
\begin{equation}
    V_\text{SR} + V_\text{LR} + V_\text{self} = \frac{1}{2}\sum_i q_i^2 \tilde{v}_\mathrm{M} + \frac{1}{2} \sum_{i\neq j} q_i q_j \tilde{v}_\mathrm{E}(\vr_i-\vr_j),
\end{equation}
where $\tilde{v}_\mathrm{E}(\vr)$ and $\tilde{v}_\mathrm{M}$ differ from Eqs.~(\ref{eq:ewald}) and (\ref{eq:madelung}) by a constant, $-\pi/\Omega\eta^2$.
As discussed in Sec.~\ref{sec:ewald}, the total energy is unchanged by addition of a constant to the Ewald potential, and this particular constant makes the potential zero on average.

Finally, what justified our neglect of $\vG=\bm{0}$ in $V_\text{LR}$?
It is not sufficient to invoke charge neutrality [$\rho(\vG=\bm{0})=0$], because the needed limit is
\begin{equation}
    \lim_{\vG\rightarrow \bm{0}} \frac{|\rho(\vG)|^2}{G^2}.
\end{equation}
For a neutral cell, the small-$\vG$ limit of $\rho(\vG)$ is
\begin{equation}
    \rho(\vG) = \sum_i q_i e^{-i\vG\cdot \vr_i} \approx -i \vG\cdot \bm{P},
\end{equation}
where $\bm{P} = \sum_i q_i \vr_i$ is the unit cell dipole moment,
giving a surface energy
\begin{equation}
    V_\text{surf} = \lim_{\vG\rightarrow \bm{0}} \frac{2\pi}{\Omega} |\hat{\vG}\cdot\bm{P}|^2.
\end{equation}
The limit depends on the direction of approach, which reflects the shape dependence of the surface energy.
Spherically averaging over all three directions yields a contribution
\begin{equation}
    V_\text{surf} \approx \frac{2\pi}{\Omega} \sum_{\alpha\beta} P_\alpha P_\beta \langle \hat{G}_\alpha \hat{G}_\beta \rangle_\text{sph} = \frac{2\pi}{3\Omega} |\bm{P}|^2,
\end{equation}
in agreement with more rigorous derivations of spherical boundary conditions.
As discussed in the main text, neglecting the $\vG=\bm{0}$ is equivalent to neglecting the surface energy. 

\bibliographystyle{aipnum4-2-arxiv}
\bibliography{tutorial}

\raggedbottom

\end{document}